\documentclass[%
reprint,
superscriptaddress,
showpacs,preprintnumbers,
 amsmath,amssymb,
 aps,
prd,
]{revtex4-1}

\usepackage{graphicx}
\usepackage{dcolumn}
\usepackage{bm}
\usepackage{bbold}
\usepackage{amssymb,amsmath}
\usepackage{hyperref}

\usepackage{color}
\usepackage{amsfonts}
\usepackage{subfigure}
\usepackage{array}

\newcommand{\Tr}{\ensuremath{\operatorname{Tr}}}

\newcolumntype{L}{>{\centering\arraybackslash}m{3cm}}

\definecolor{bjcol}{rgb}{1,.44,0.13}

\definecolor{blue}{rgb}{0,0,1}

\definecolor{green}{rgb}{0,1,0}

\definecolor{red}{rgb}{1,0,0}

\definecolor{gray}{rgb}{.5,.5,.5}

\definecolor{darkgreen}{rgb}{.0,.5,.0}

\def\Fig#1{Fig.~\ref{#1}} \def\Tab#1{Tab.~\ref{#1}}
 \def\Tab#1{Tab.~\ref{#1}}

\def\Eq#1{Eq.~(\ref{#1})}

\def\eqref#1{(\ref{#1})}

\def\sec#1{Sec.~\ref{#1}}
\def\app#1{Appendix~\ref{#1}}

\def\lA0{{\langle A_0 \rangle}}
\def\bA0{{\bar{A}_0}}

\def\0#1#2{\frac{#1}{#2}}

\graphicspath{{./figures/}{./}}

\begin{document}

\preprint{}

\title{Baryon number fluctuations in the 2+1 flavor low energy effective model
}

\author{Rui Wen}
\affiliation{School of Physics, Dalian University of Technology, Dalian, 116024,
  P.R. China}

\author{Chuang Huang}
\affiliation{School of Physics, Dalian University of Technology, Dalian, 116024,
  P.R. China}

\author{Wei-jie Fu}
\email{wjfu@dlut.edu.cn}
\affiliation{School of Physics, Dalian University of Technology, Dalian, 116024,
  P.R. China}


\begin{abstract}

High order cumulants of the baryon number distribution are calculated in a 2+1 flavor low energy effective model. Quantum fluctuations are encoded through the functional renormalization group approach. The chiral and deconfinement phase transitions are investigated at finite temperature and baryon chemical potential. The equation of state for the QCD matter, fluctuations of the baryon number and the strangeness up to sixth order are calculated, and one finds the results are consistent with those from lattice QCD.

\end{abstract}

\maketitle


\section{Introduction}
\label{sec:intro}

Studying the phase structure of the hot and/or dense QCD matter has attracted lots of attentions in recent years. In the QCD phase diagram spanned by the temperature and the chemical potential, the critical end point (CEP), which separates the first-order phase transition at high chemical potential from the continuous crossover at high temperature, plays a central role due to its uniqueness and significance. Unfortunately, location, and even existence, of the CEP is hitherto unclear. Lots of efforts, however, have been made to be aimed at unravelling the mysterious veil. In the experiments significant progress has been made in the Beam Energy Scan (BES) Program at the Relativistic Heavy Ion Collider (RHIC) \cite{Adamczyk:2013dal,Adamczyk:2014fia,Adamczyk:2017iwn}, see a review in \cite{Luo:2017faz}. 

In the meantime theoretical studies of the phase structure of dense QCD matter, as well as confrontation of theoretical calculations with experimental data, are currently ongoing research frontier. In recent year, the QCD equation of state (EoS), fluctuations and correlations of conserved charges, etc. have been computed and studied systematically in lattice QCD at finite chemical potential, see e.g. \cite{Bazavov:2012vg,Bazavov:2014pvz,Bazavov:2017dus,Bazavov:2017tot,Borsanyi:2013bia,Borsanyi:2013hza,Borsanyi:2014ewa,Borsanyi:2018grb}. In particular, freeze-out parameters in heavy ion collisions are already accessible through comparison between theoretical calculations and experiments \cite{Bazavov:2012vg,Borsanyi:2013hza,Borsanyi:2014ewa}.  Functional continuum field approaches besides lattice QCD also provide us with a wealth of knowledge about the QCD phase structure \cite{Braun:2009gm,Qin:2010nq,Fischer:2013eca} and the properties of QCD matter \cite{Christiansen:2014ypa,Cyrol:2018xeq}, especially in the regime which can not be reached by lattice simulations. 

Of observables related to the search of CEP, fluctuations of conserved charges, e.g. the net baryon or proton, are of extreme importance. Baryon number fluctuations have been studied in the two flavor low energy effective theory within the functional renormalization group (FRG) approach in \cite{Skokov:2010sf,Skokov:2011yb,Fu:2015naa}, and consistent results have been observed with the improvement of the truncation for the FRG \cite{Fu:2016tey}. In the FRG approach quantum fluctuations are encoded through the running of flows, cf. reviews \cite{Berges:2000ew,Pawlowski:2005xe,Pawlowski:2010ht,Braun:2011pp}.  As a nonperturbative continuum field approach, FRG has been successfully applied in first-principle QCD calculations, and significant progress has been made in recent year, see, e.g., \cite{Mitter:2014wpa,Braun:2014ata,Cyrol:2016tym,Cyrol:2017ewj,Cyrol:2017qkl} for more details. Moreover, it has also been widely used in low energy effective theories, see e.g., \cite{Schaefer:2004en,Schaefer:2006ds,Nakano:2009ps,Herbst:2010rf,Pawlowski:2014zaa,Helmboldt:2014iya,Wang:2015bky,Fu:2017vvg,Zhang:2017icm,Almasi:2017bhq}.

Given the importance of the baryon number fluctuations in the phenomenology of CEP searching in the experiments \cite{Stephanov:1999zu,Xin:2014ela,Fu:2015amv,Li:2017ple,Li:2018ygx}, in this work we would like to extend relevant studies in Refs. \cite{Fu:2015naa,Fu:2015amv,Fu:2016tey} from two to 2+1 flavors, based on the 2+1 flavor Polyakov-loop improved quark-meson (PQM) effective model, see, e.g. \cite{Schaefer:2007pw,Schaefer:2009ui,Schaefer:2011ex,Mitter:2013fxa,Herbst:2013ufa,Rennecke:2016tkm,Fu:2018qsk,Fu:2018swz} for more details about the model, and Refs. \cite{Fu:2007xc,Fu:2009wy,Fu:2010ay} for other relevant 2+1 flavor low energy effective models of the same class. Note that including the strangeness is nontrivial, since it introduce the degrees of freedom of not only the strange quarks but also open-strange mesons, such as kaon mesons.

From another point of view, the 2+1 flavor low energy effective model is the hadronic sector of the 2+1 flavor rebosonized QCD, and is readily embedded into the glue sector within the FRG approach, see \cite{Mitter:2014wpa,Braun:2014ata} for the two flavor rebosonized QCD. Therefore, the 2+1 flavor low energy effective model investigated in this work is also aimed to facilitate for the construction of the 2+1 flavor rebosonized QCD in the near future. Therefore, we would like to describe the theoretical framework in more detail. The QCD thermodynamics, such as the pressure and the EoS at finite temperature and baryon chemical potential, will be investigated, and then we will compare our calculated results of the baryon number fluctuations with those from lattice simulations. 

This paper is organized as follows. In Sec.~\ref{sec:PQM} we describe the 2+1 flavor low energy effective model. The FRG and the flow equation for the effective potential are given in Sec.~\ref{sec:FRG}. We will discuss the thermodynamics and the baryon number fluctuations in Sec.~\ref{sec:thermo}. Numerical results are presented in Sec.~\ref{sec:Num}, and then a summary and an outlook are presented in Sec.~\ref{sec:sum}. Furthermore, more technical details about the flow equation of the effective potential, the glue potential, and the numerical setup are given in \app{app:flowV}, \app{app:gluepot}, and \app{app:num}, respectively.


\section{2+1 flavor low energy effective model}
\label{sec:PQM}

In this work we adopt the 2+1 flavor Polyakov-loop improved quark-meson model \cite{Mitter:2013fxa,Herbst:2013ufa,Rennecke:2016tkm,Fu:2018qsk}, and quantum fluctuations of different scales, as well as the thermal and density fluctuations, are encoded successively through the evolution of the renormalization group (RG) scale dependent effective action $\Gamma_{k}$, which in the Euclidean formalism reads
\begin{align}
	\Gamma_{k}[\Phi]=&\int_x \bigg\{ \bar{q} [\gamma_\mu \partial_\mu-\gamma_0(\hat\mu+igA_0)]q+h\,\bar{q} \,\Sigma_5 q\nonumber\\[2ex]
&+\text{tr}(\bar D_\mu \Sigma \cdot \bar D_\mu\Sigma^\dagger)+\tilde{U}_{k}(\Sigma)+V_{\text{\tiny{glue}}}(L, \bar L)\bigg\}\,,\label{eq:action}
\end{align}
with $\Phi$ denoting all the field dependence, where we have used a shorthand notation  $\int_{x}=\int_0^{1/T}d x_0 \int d^3 x$ with the temperature $T$. The quark chemical potential $\hat{\mu}$ is a diagonal matrix in the flavor space, i.e., $\hat{\mu}=\mathrm{diag}(\mu_u,\mu_d,\mu_s)$ with $\mu_u=\mu_B/3+2\mu_Q/3$, $\mu_d=\mu_B/3-\mu_Q/3$, and $\mu_s=\mu_B/3-\mu_Q/3-\mu_S$, where $\mu_B$, $\mu_Q$, and $\mu_S$ are the chemical potentials related to the baryon number, electric charge, and the strangeness, respectively. The covariant derivative of meson fields reads
\begin{align}
\bar D_\mu \Sigma=\partial_\mu +\delta_{\mu 0}[ \hat \mu ,\Sigma]\,.
\end{align}
Note that the mesons do not carry the baryon chemical potential, but they could have $\mu_Q$ and $\mu_S$, see, e.g., \cite{Fu:2018qsk} for details. Note also that the local potential approximation (LPA) to the effective action is implicitly implied in \Eq{eq:action}, viz. only the effective potential $\tilde{U}_{k}(\Sigma)$ is dependent on the RG scale $k$. For more discussions about truncations beyond LPA, taking into account, for instance, the nontrivial dispersion relations and the running Yukawa coupling, etc., see e.g. \cite{Rennecke:2016tkm}.

The temporal gluon background field $A_0$ in \Eq{eq:action} is related to the color confinement and its phase transition, which could be formulated as the Polyakov loops for convenience, to wit,
\begin{align}
  L(\bm{x})=\frac{1}{N_c} \left\langle \Tr\, {\cal P}(\bm x)\right\rangle \,,\quad  \bar L (\bm{x})=\frac{1}{N_c} \langle
  \Tr\,{\cal P}^{\dagger}(\bm x)\rangle \,,\label{eq:Lloop}
\end{align}
with 
\begin{align}
  {\cal P}(\bm x)= \mathcal{P}\exp\Big(ig\int_0^{\beta}d\tau A_0(\bm{x},\tau)\Big)\,,\label{eq:Ploop}
\end{align}
where $\mathcal{P}$ on the r.h.s. denotes the path ordering. The Polyakov loop, from the viewpoint of statistics,  can be regarded as the order parameter of the Z(3) symmetry for the deconfinement phase transition. The dynamics of the Polyakov loop is governed by the glue potential, also called as the Polyakov loop potential $V_{\text{\tiny{glue}}}(L, \bar L)$ in \Eq{eq:action}. Usually the glue potential and its dependence on the external parameters, such as the temperature, can be parameterized by employing first-principle QCD calculations, for instance the lattice computation \cite{Lo:2013hla} and the FRG \cite{Herbst:2015ona}. Then, the parameterized glue potentials are applied in low energy effective theories, for more relevant discussions, see, e.g. the review article \cite{Fukushima:2017csk} and reference therein.

In the effective action \Eq{eq:action}, quarks are coupled to the scalar and pseudoscalar meson nonets via a chirally symmetric Yukawa term with
\begin{align}
  \Sigma_5&=T^a(\sigma^a+i \gamma_5\pi^a)\,, \quad a=0,\,1,...,8\,,\label{}
\end{align}
where $T^a$ are the generators of the flavor $U(N_f)$ group, which for the superscript $a=1,...,8$ can be represented by the Gell-Mann matrices, i.e., $T^a=\lambda^a/2$, and $T^{0}=\frac{1}{\sqrt{2N_{f}}}\mathbb{1}_{N_{f}\times N_{f}}$. The kinetic term for the mesons in \Eq{eq:action} is formulated in terms of the adjoint representation of $U(N_f)$, which reads
\begin{align}
  \Sigma&=T^a(\sigma^a+i \pi^a)\,. \label{}
\end{align}

In \Eq{eq:action} $\tilde{U}_{k}(\Sigma)$ is the meson effective potential, which consists of several parts as follow
\begin{align}
  \tilde{U}_{k}(\Sigma)&=U_k(\rho_1,\tilde{\rho}_2)-c_A \xi-j_L\sigma_L-j_S\sigma_S\,, \label{eq:tildeU}
\end{align}
with $U_k(\rho_1,\tilde{\rho}_2)$ on the r.h.s. being an arbitrary function of $\rho_1$ and $\tilde{\rho}_2$, where $\rho_1$ and $\tilde{\rho}_2$ are invariant under the transformation of $SU_{\text{V}}(3)\times SU_{\text{A}}(3)\times U_{\text{V}}(1)\times U_{\text{A}}(1)$. Therefore, $U_k(\rho_1,\tilde{\rho}_2)$ has the maximal symmetry of flavors, and the invariants $\rho_1$ and $\tilde{\rho}_2$ are defined as 
\begin{align}
  \rho_1&=\text{tr}(\Sigma \cdot \Sigma^\dagger)\,, \label{eq:rho1}\\[2ex]
  \tilde{\rho}_2&=\text{tr}\Big(\Sigma \cdot \Sigma^\dagger-\frac{1}{3}\,\rho_1\,\mathbb{1}_{3\times 3}\Big)^2 \,.\label{eq:rho2}
\end{align}
$\xi$ in \Eq{eq:tildeU} is the Kobayashi-Maskawa-'t Hooft determinant which reads
\begin{align}
  \xi&=\det(\Sigma)+\det(\Sigma^\dagger)\,, \label{}
\end{align}
and the relevant term breaks the $U_{\text{A}}(1)$ symmetry, which stems from the axial anomaly due to quantum fluctuations of QCD. The coefficient $c_A$ is scale independent. The last two terms linear in the sigma fields in \Eq{eq:tildeU} break the chiral symmetry explicitly, which results directly  in mass acquirement for the Goldstone bosons, such as pions and kaons, etc. In \Eq{eq:tildeU} we have employed the light-strange basis, which is related to the singlet-octet one through the relation as follows
\begin{align}
  \begin{pmatrix} \phi_L \\ \phi_S \end{pmatrix}
  &=\frac{1}{\sqrt{3}}\begin{pmatrix} 1 & \sqrt{2}\\ -\sqrt{2} & 1 \end{pmatrix}
  \begin{pmatrix}\phi_8\\ \phi_0 \end{pmatrix}\,, \label{}
\end{align}
with $\phi$ denoting scalar and pseudoscalar mesons collectively. Apparently, the strength of explicit breaking of the chiral symmetry, in another word, how massive the Goldstone bosons are, is connected to the magnitude of the coefficients $j_L$ and $j_S$ in \Eq{eq:tildeU}, which in this work are regarded as the RG scale independent parameters to be determined in the following. The constituent quark masses for the light and strange quarks are given by
\begin{align}
  m_l&=\frac{h}{2}\bar \sigma_L, \quad m_s=\frac{h}{\sqrt{2}}\bar \sigma_S\,,\label{}
\end{align}
respectively, and the pion and kaon decay constants read \cite{Lenaghan:2000ey}
\begin{align}
  f_\pi=\bar \sigma_L, \quad f_K=\frac{\bar \sigma_L+\sqrt{2}\,\bar \sigma_S}{2}\,,\label{eq:fpiK}
\end{align}
where $\bar \sigma_L$ and $\bar \sigma_S$ denotes the expected values of $\sigma_L$ and $\sigma_S$ fields. The meson mass squares are obtained by diagonalizing the Hessian matrix of the effective potential, which reads
\begin{align}
  H_{ij}=\frac{\partial^2 \tilde{U}_{k}}{\partial \phi_i \partial \phi_j}\,.\label{}
\end{align}
The Hessian matrix is block diagonal in the scalar and pseudoscalar channels, and in each block the only nonvanishing nondiagonal element is $H_{80}$ or $H_{08}$, which resulting in the mixing of mesons between the octet and singlet. We will not go into the details in this work, and for more relevant discussions as well as explicit expressions for the meson masses, see, e.g. \cite{Rennecke:2016tkm}.


\section{Quantum fluctuations within FRG}
\label{sec:FRG}

%
\begin{figure*}[t]
\includegraphics[width=1.\textwidth]{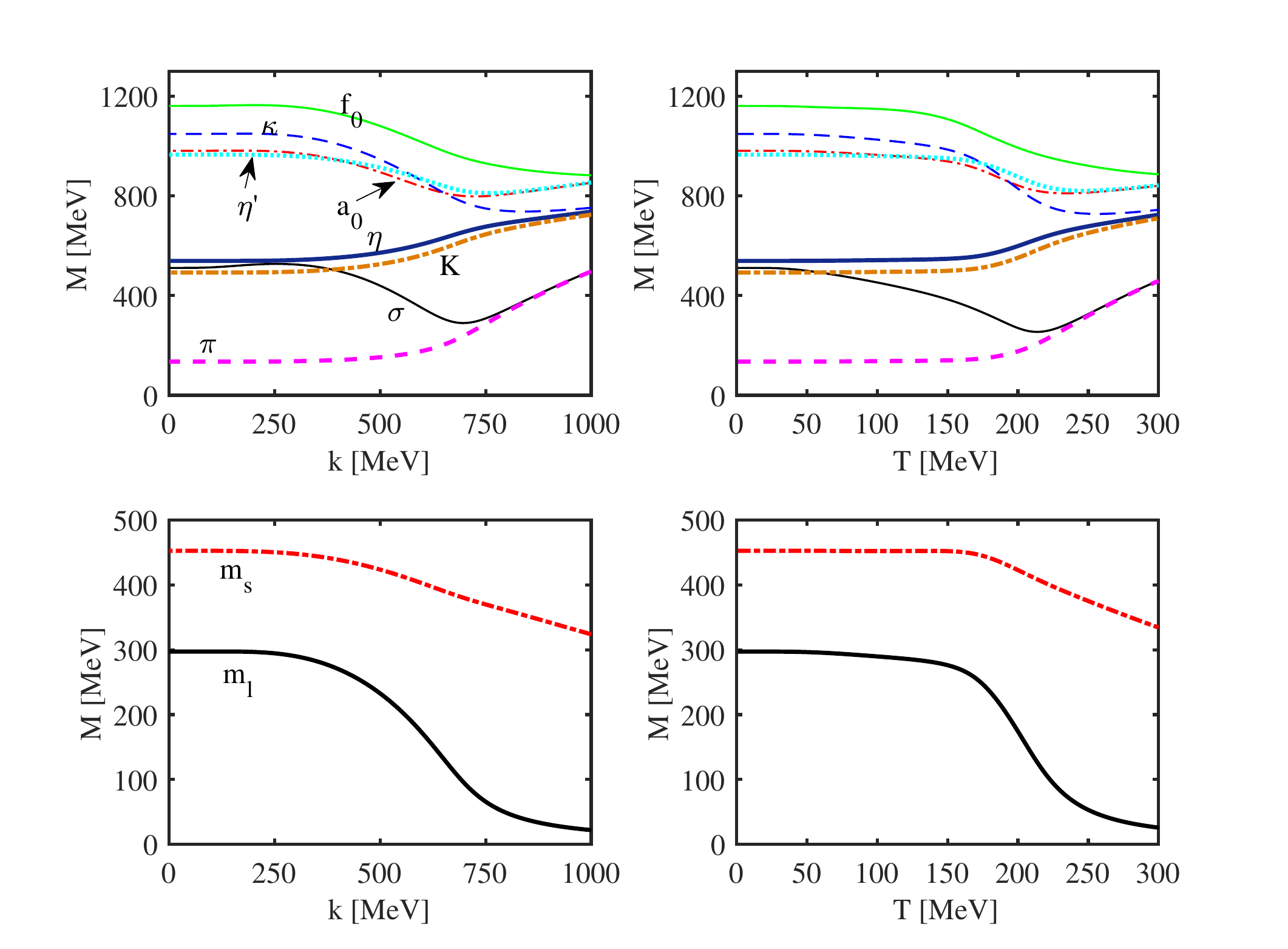}
\caption{Left: running of the meson (top) and quark (bottom) masses with the RG scale $k$ at vacuum. Right:  meson (top) and quark (bottom) masses as functions of the temperature at vanishing chemical potential. Calculations at finite temperature are performed by employing the glue potential $V_{\text{glue-Haar}}$ in \Eq{eq:GluepHaar} with $T_c^\text{\tiny{glue}}=270$ MeV and $\alpha=0.52$.}\label{fig:mass}
\end{figure*}
%

In the RG scale dependent effective action $\Gamma_{k}$ in \Eq{eq:action}, quantum fluctuations of wavelength $\gtrsim 1/k$ are suppressed, and the scale $k$ is in fact an infrared (IR) cutoff scale. Therefore, by lowering the IR cutoff scale toward the IR limit, i.e., $k\rightarrow 0$, one arrives at a full quantum effective action $\Gamma_{k\rightarrow 0}$. As a consequence, one also has to specify the initial ultraviolet (UV) scale where the evolution begins. An ideal choice is a scale deep in the perturbative regime, for instance the mass scale of the $Z$ boson, which, however, entails the embedding of the glue dynamics. The dynamics of glue part, including the gluon and ghost in the Landau gauge for example, is indispensable to the evolution of $\Gamma_{k}$, when $k$ is above $\sim$ 1 GeV, thus the effective action in \Eq{eq:action} is insufficient at high scale. For more discussions about the first-principle FRG QCD calculations and recent progresses whereof, see e.g. \cite{Pawlowski:2014aha,Mitter:2014wpa,Braun:2014ata,Cyrol:2016tym,Cyrol:2017ewj,Cyrol:2017qkl,Cyrol:2018xeq}. It has been known that, with the evolution of the RG scale $k$ from the UV to IR regime, the glue dynamics decouples from other degrees of freedom when $k$ is reduced to about 1 GeV, since a mass gap of the gluon field develops therein \cite{Braun:2014ata,Cyrol:2016tym,Cyrol:2017ewj,Cyrol:2017qkl}. Therefore, one can start the evolution of flow equations at a scale $\Lambda\sim$ 1 GeV, safely neglecting the glue part, and the quantities at the initial scale are the parameters of the low energy effective model, which are needed to be determined.

The flow equation for the effective action in \Eq{eq:action}, i.e., the Wetterich equation \cite{Wetterich:1992yh}, is given by
\begin{align}
  \partial_{t}\Gamma_{k}&=-\mathrm{Tr}\big(G^{q\bar q}_{k}\partial_{t} R^{q}_{k}\big)+\frac{1}{2}\mathrm{Tr}\big(G^{\phi\phi}_{k}\partial_{t} R^{\phi}_{k}\big)\,,
 \label{eq:WetterichEqPQM}
\end{align}
with the RG time being $t=\ln (k/\Lambda)$, where $\Lambda$ is the initial evolution scale as mentioned above, and it is also called as the UV cutoff. The two terms on the r.h.s. of \Eq{eq:WetterichEqPQM} corresponds to contributions from the quarks and mesons, respectively, and in our case there are three flavor quarks, meson nonets in the scalar and pseudoscalar channels. In \Eq{eq:WetterichEqPQM} $G_k$'s are the propagators of quarks and mosons, and the IR regulators $R_k$'s suppress quantum fluctuations of wavelengths $\gtrsim 1/k$. 

With the truncation of LPA, the Wetterich equation in \Eq{eq:WetterichEqPQM} is equivalent to the flow equation for the effective potential, since all the $k$-dependence is encoded in such potential, which reads
\begin{align}
  &\partial_{t}U_k(\rho_1,\tilde{\rho}_2)\nonumber\\[2ex]
  =&\frac{k^4}{4\pi^2}\bigg\{l_{0}^{(B)}(\bar m_{\pi,k}^{2};T,0)+2\,l_{0}^{(B)}(\bar m_{\pi,k}^{2};T,\mu_u-\mu_d)\nonumber\\[2ex]
&+2\,l_{0}^{(B)}(\bar m_{K,k}^{2};T,\mu_u-\mu_s)+2\,l_{0}^{(B)}(\bar m_{K,k}^{2};T,\mu_d-\mu_s)\nonumber\\[2ex]
&+l_{0}^{(B)}(\bar m_{\eta,k}^{2};T,0)+l_{0}^{(B)}(\bar m_{\eta^\prime,k}^{2};T,0)\nonumber\\[2ex]
&+l_{0}^{(B)}(\bar m_{a_0,k}^{2};T,0)+2\,l_{0}^{(B)}(\bar m_{a_0,k}^{2};T,\mu_u-\mu_d)\nonumber\\[2ex]
&+2\,l_{0}^{(B)}(\bar m_{\kappa,k}^{2};T,\mu_u-\mu_s)+2\,l_{0}^{(B)}(\bar m_{\kappa,k}^{2};T,\mu_d-\mu_s)\nonumber\\[2ex]
&+l_{0}^{(B)}(\bar m_{f_0,k}^{2};T,0)+l_{0}^{(B)}(\bar m_{\sigma,k}^{2};T,0)\nonumber\\[2ex]
&-4N_{c}\Big[l_{0}^{(F)}(\bar m_{l,k}^{2};T,\mu_u)+l_{0}^{(F)}(\bar m_{l,k}^{2};T,\mu_d)\nonumber\\[2ex]
&+l_{0}^{(F)}(\bar m_{s,k}^{2};T,\mu_s)\Big]\bigg\}\,,\label{eq:dtUk}
\end{align}
with the dimensionless mass square $\bar m_{i, k}^{2}\equiv m_{i, k}^{2}/k^2$ and the threshold functions $l_{0}^{(B/F)}$ given in \app{app:flowV},  where all the scalar and pseudoscalar mesons in the nonets are shown explicitly, and the number 2 in front of $l_{0}^{(B)}$'s denote the degeneracy of the meson and its charge conjugate. The dependence of the threshold functions on the temperature and chemical potential is also shown in \Eq{eq:dtUk}. Note that the baryon chemical potential does not enter into $l_{0}^{(B)}$'s, since the mesons do not carry the baryon number. However, this is not the case for other chemical potentials, such as those for the electric charge and strangeness, which can be carried by a meson, for more relevant discussions, see, e.g. \cite{Fu:2018qsk}. In this work we employ the Taylor expansion around the physical point to solve the flow equation for the effective potential in \Eq{eq:dtUk}, which is presented in detail in \app{app:flowV}.


\section{Thermodynamics and the baryon number fluctuations}
\label{sec:thermo}

%
\begin{figure}[t]
\includegraphics[width=0.5\textwidth]{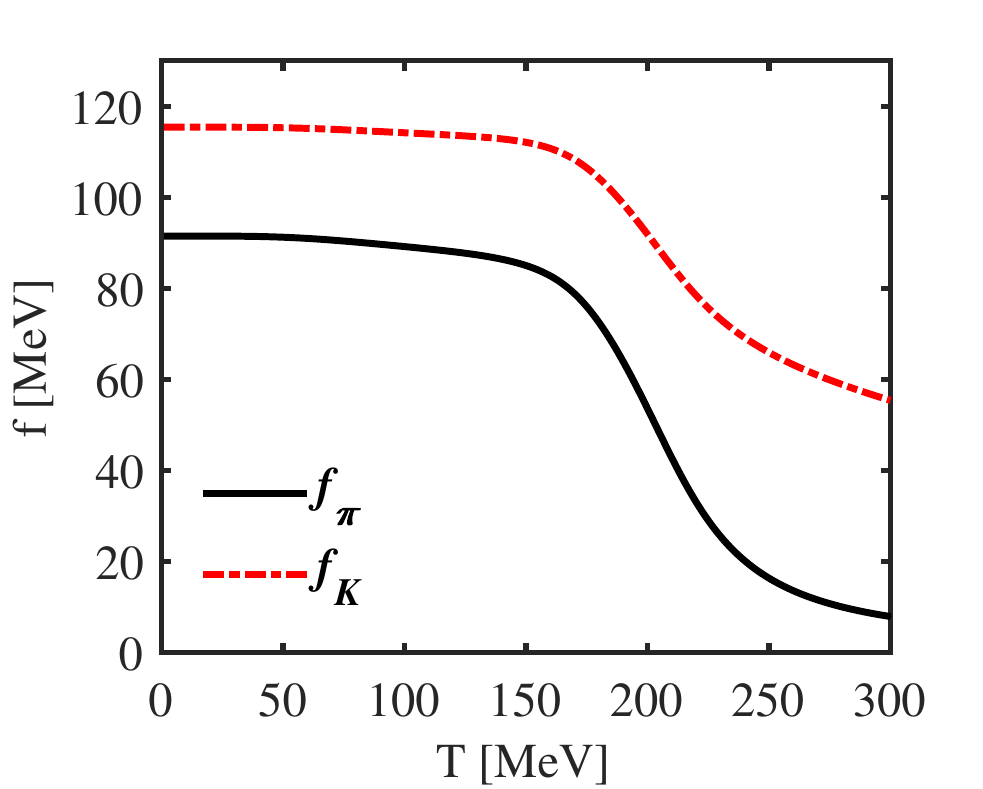}
\caption{Pion and kaon decay constants as functions of the temperature at vanishing chemical potential. The glue potential $V_{\text{glue-Haar}}$ in \Eq{eq:GluepHaar} with $T_c^\text{\tiny{glue}}=270$ MeV and $\alpha=0.52$ is employed.}\label{fig:fpifK}
\end{figure}
%

The thermodynamical potential density is connected to the effective action $\Gamma_{k}$ in \Eq{eq:action}, more exactly its IR limit with $k=0$, through the relation as follows
\begin{align}
  \Omega[T, \mu]=\frac{T}{V}\left(\Gamma_{k=0}[\bar \Phi]\Big\vert_{T,\mu}-\Gamma_{k=0}[\bar \Phi]\Big\vert_{T=\mu=0}\right)\,,\label{eq:Omega}
\end{align}
where $\bar \Phi$ are the physical values of fields, i.e., the solutions of their equations of motion, and $V$ is the volume of the system. Upon inserting \Eq{eq:action} into \Eq{eq:Omega} and considering the expected values of all the fields, one arrives at
\begin{align}
  \Omega[T, \mu]=\tilde{U}_{k=0}(\bar{\sigma}_L,\bar{\sigma}_S)+V_{\text{\tiny{glue}}}(L, \bar L)\,,\label{}
\end{align}
where it is assumed that $\Omega$ has been normalized to be vanishing at vacuum as \Eq{eq:Omega}. In order to investigate the dependence of our calculated results on the parameterization scheme for the glue potential, in this work we employ two different glue potentials, which are commonly used in literatures, i.e., the polynomial potential in \Eq{eq:Glueppoly} and that with the Haar measure in \Eq{eq:GluepHaar}, respectively. More detailed descriptions and discussions about these two potentials are deferred to \app{app:gluepot}.

With the thermodynamical potential in \Eq{eq:Omega} in hands, one can obtain other equilibrium thermodynamical quantities, such as the pressure and the entropy density:
\begin{align}
  p&=-\Omega[T, \mu]\,, \quad \text{and} \quad s=\frac{\partial p}{\partial T}\,.\label{}
\end{align}
The equation of state for the QCD matter is well describe by the interaction measure, i.e., the trace anomaly which reads
\begin{align}
  \Delta&=\epsilon-3p\,,\label{}
\end{align}
with the energy density $\epsilon$.
 
The $n$-th order cumulants of the net baryon number $N_B$ distributions are given by
\begin{align}
  \langle (\delta N_B)^n \rangle&=\sum_{N_B=-\infty}^{\infty}(\delta N_B)^n P(N_B)\,,\label{eq:dNBaver}
\end{align}
with $\delta N_B=N_B-\langle N_B\rangle$, where $P(N_B)$ is the probability distribution of $N_B$, and $\langle N_B\rangle$ its mean value. Therefore, theoretical calculations of the cumulants are feasible, if the probability distribution $P(N_B)$ is obtained. In fact, $P(N_B)$ can be obtained in the same theoretical framework by resorting to the canonical partition function with an imaginary chemical potential, see e.g. \cite{Morita:2013tu,Sun:2018ozp} for more details. It is also found that in the low energy effective model with the Polyakov loop, the cumulants obtained in \Eq{eq:dNBaver} agree with those obtained from the generalized susceptibilities in \Eq{eq:suscept} in the following \cite{Sun:2018ozp}. The more commonly used generalized susceptibilities, viz., computing the $n$-th order derivative of the pressure w.r.t. the baryon chemical potential, are given by
\begin{align}
   \chi_n^{\text{B}}&=\frac{\partial^n}{\partial (\mu_B/T)^n}\frac{p}{T^4}\,.\label{eq:suscept}
\end{align}
They are related to the cumulants of the baryon number distribution through relations, for instance up to the fourth order, as follow
\begin{subequations}\label{eq:chi1to4}
\begin{align}
  \chi_1^{\text{B}}&=\frac{1}{VT^3}\langle N_B \rangle\,,\\[2ex]
  \chi_2^{\text{B}}&=\frac{1}{VT^3}\langle(\delta N_B)^2\rangle\,,\\[2ex]
  \chi_3^{\text{B}}&=\frac{1}{VT^3}\langle(\delta N_B)^3\rangle\,,\\[2ex]
  \chi_4^{\text{B}}&=\frac{1}{VT^3}\Big(\langle(\delta N_B)^4\rangle-3\langle(\delta N_B)^2\rangle^2\Big)\,,
\end{align}
\end{subequations}
In the experimental measurements, the mean value $M$, variance $\sigma^2$, skewness $S$, and the kurtosis $\kappa$ of the net proton or baryon number distribution are usually used, which are connected to the generalized susceptibilities in \Eq{eq:chi1to4} by
\begin{align}
  M&=VT^3\chi_1^{\text{B}}\,,\qquad \sigma^2=VT^3\chi_2^{\text{B}}\,,\nonumber\\[2ex]
  S&=\frac{\chi_3^{\text{B}}}{\chi_2^{\text{B}}\sigma}\,,\qquad \hspace{0.55cm}\kappa=\frac{\chi_4^{\text{B}}}{\chi_2^{\text{B}}\sigma^2}\,.
\end{align}
Furthermore, in this work we also calculate the sixth-order susceptibility at vanishing chemical potential, which reads
\begin{align}
    \chi_6^{\text{B}}&=\frac{1}{VT^3}\Big(\langle(\delta N_B)^6\rangle-15\langle(\delta N_B)^4\rangle\langle(\delta N_B)^2\rangle\nonumber\\[2ex]
&-10\langle(\delta N_B)^3\rangle^2+30\langle(\delta N_B)^2\rangle^3\Big)\,.\label{}
\end{align}


\section{Numerical results}
\label{sec:Num}

%
\begin{figure*}[t]
\includegraphics[width=1.\textwidth]{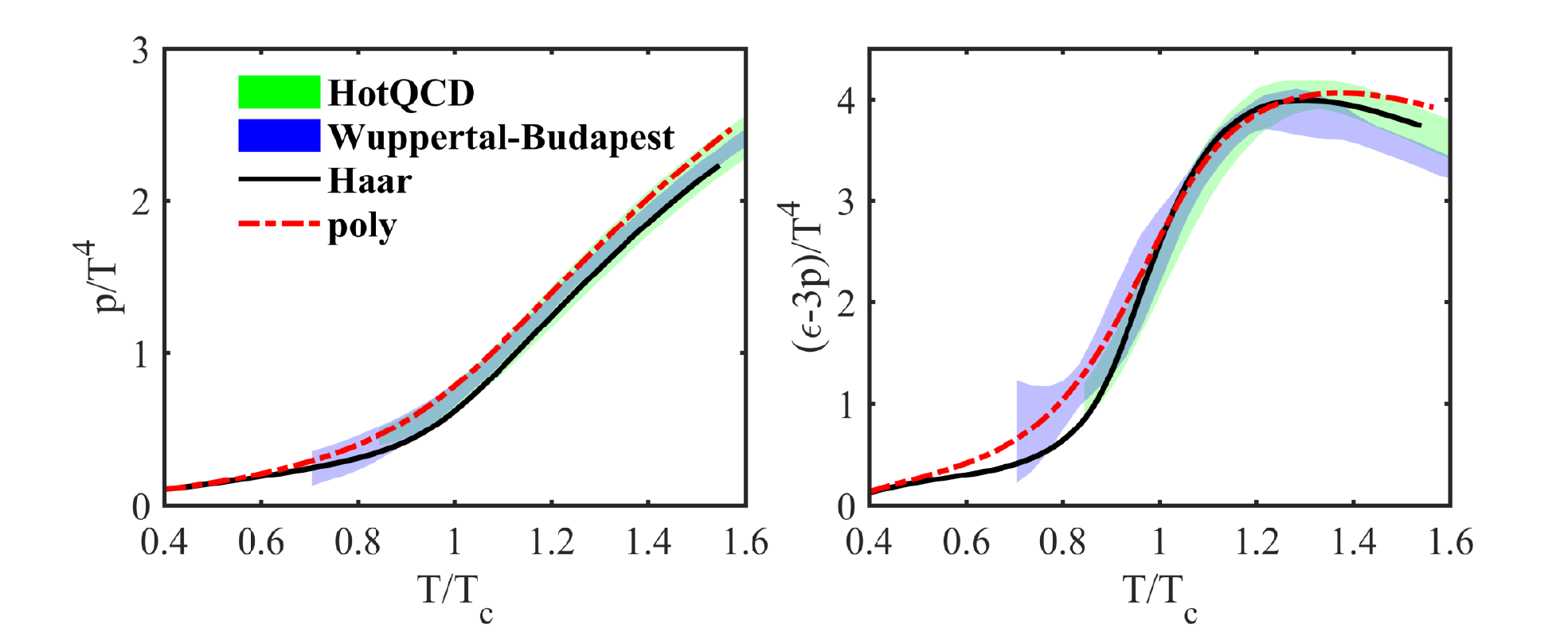}
\caption{Pressure (left panel) and trace anomaly (right panel) as functions of the temperature in unit of $T_c$ with $\mu_B=0$, where the solid and dashed lines correspond to the results obtained from the Haar glue potential $V_{\text{glue-Haar}}$ in \Eq{eq:GluepHaar} and the polynomial potential $V_{\text{{glue-poly}}}$ in \Eq{eq:Glueppoly}, respectively. Lattice results by HotQCD Collaboration \cite{Bazavov:2014pvz} and by Wuppertal-Budapest Collaboration \cite{Borsanyi:2013bia} are also presented for comparison.}\label{fig:PT4Tr}
\end{figure*}
%

%
\begin{figure*}[t]
\includegraphics[width=1.\textwidth]{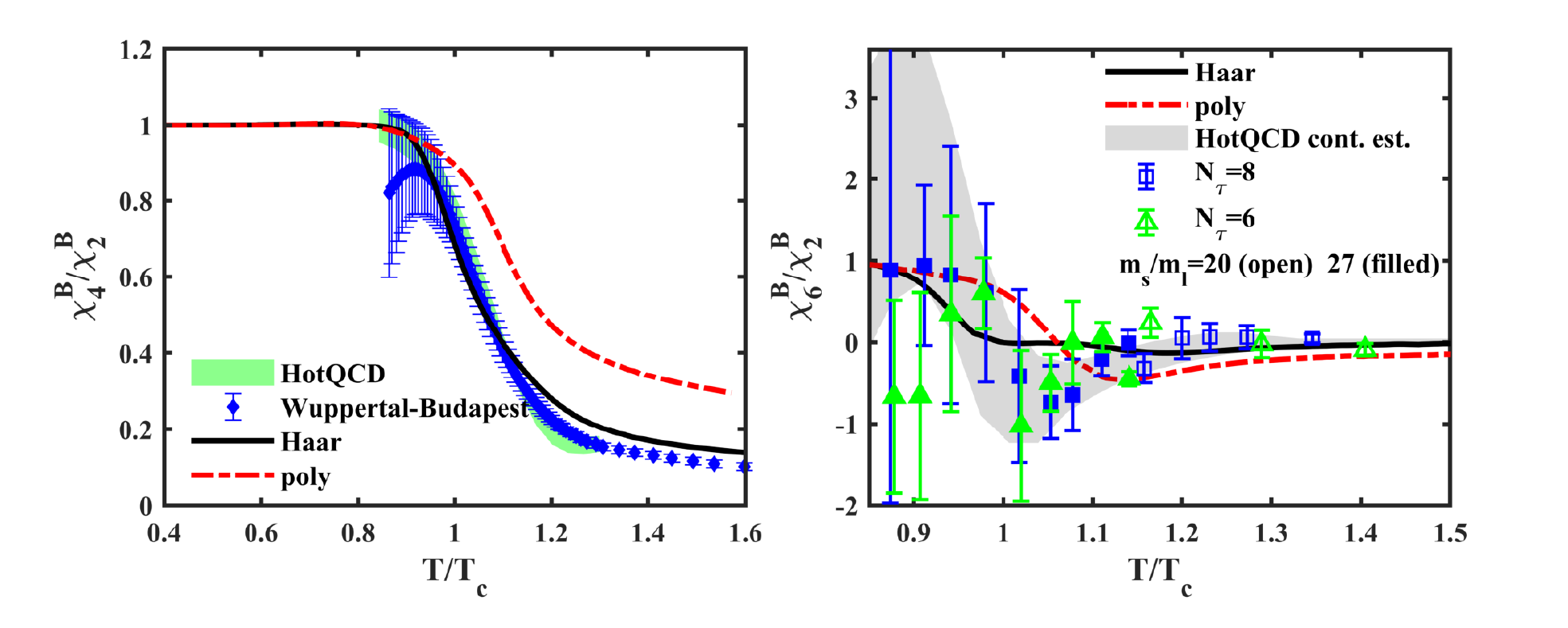}
\caption{Ratio $\chi_4^{\text{B}}/\chi_2^{\text{B}}$ (left panel) and $\chi_6^{\text{B}}/\chi_2^{\text{B}}$ (right panel) as functions of the temperature in unit of $T_c$ with $\mu_B=0$. We compare our calculated results with lattice QCD simulations by HotQCD Collaboration \cite{Bazavov:2017dus} and by Wuppertal-Budapest Collaboration \cite{Borsanyi:2013hza} in the left panel, and that by HotQCD Collaboration \cite{Bazavov:2017dus} in the right panel.}\label{fig:R4262}
\end{figure*}
%

%
\begin{figure*}[t]
\includegraphics[width=1.\textwidth]{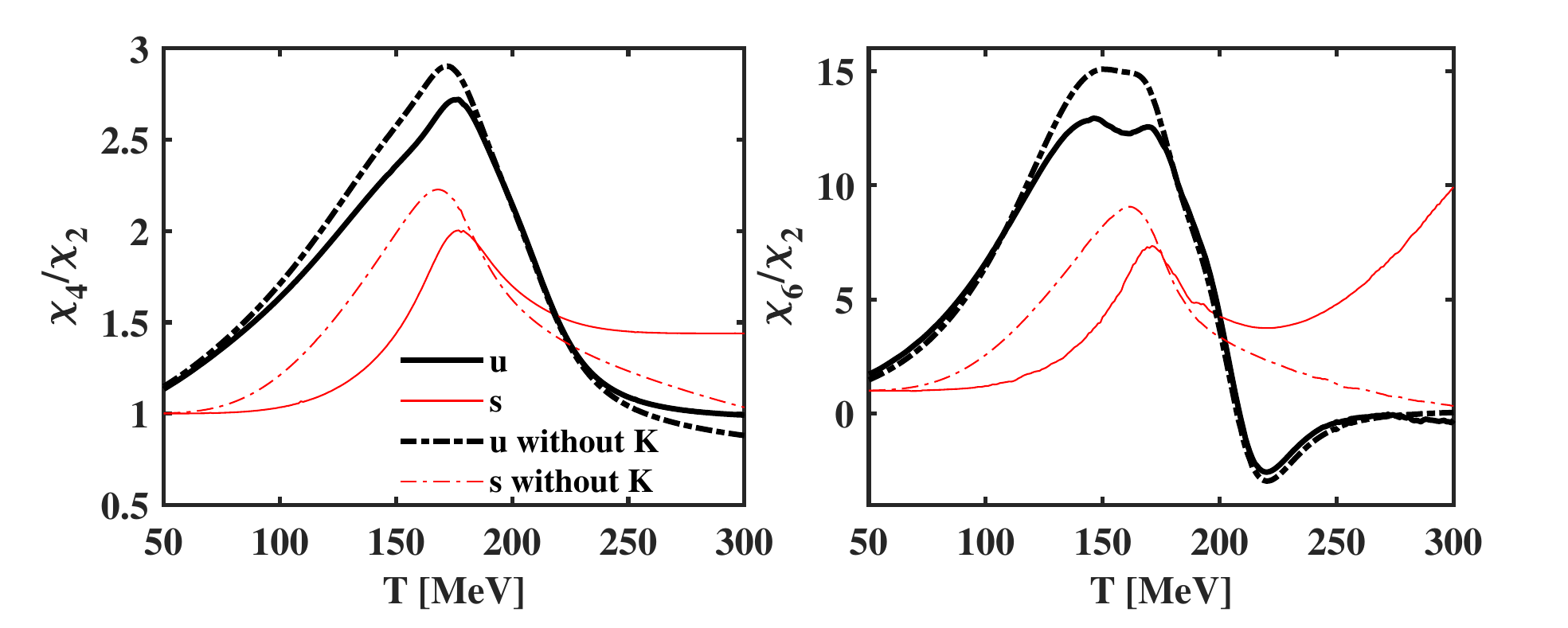}
\caption{$\chi_4/\chi_2$ and $\chi_6/\chi_2$ for $u$ and $s$ quarks as functions of the temperature at $\mu_B=0$ with and without $K/\kappa$ mesonic loops, where the Haar potential is employed.}\label{fig:R4262usK}
\end{figure*}
%

%
\begin{figure}[t]
\includegraphics[width=0.5\textwidth]{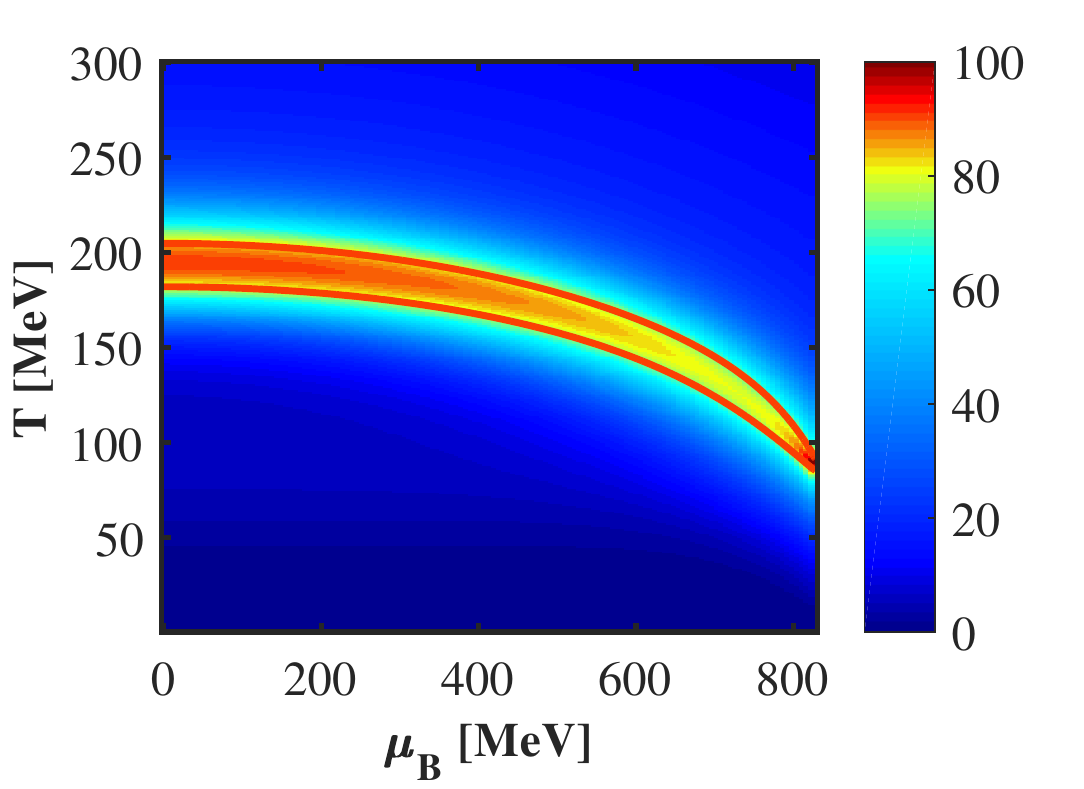}
\caption{Phase diagram of the 2+1 flavor low energy effective model in the plane of $T$ and $\mu_B$, where the Haar glue potential with $T_c^\text{\tiny{glue}}=270$ MeV and $\alpha=0.52$ is used. Color in the diagram stands for the value of $\big\vert\partial \rho_1(T,\mu_B)/\partial T \big\vert$ with $\rho_1$ given in \Eq{eq:rho1bar}. The two solid lines are determined by 90\% peak height of this derivative at each value of $\mu_B$.}\label{fig:drhodT}
\end{figure}
%

%
\begin{figure*}[t]
\includegraphics[width=1.\textwidth]{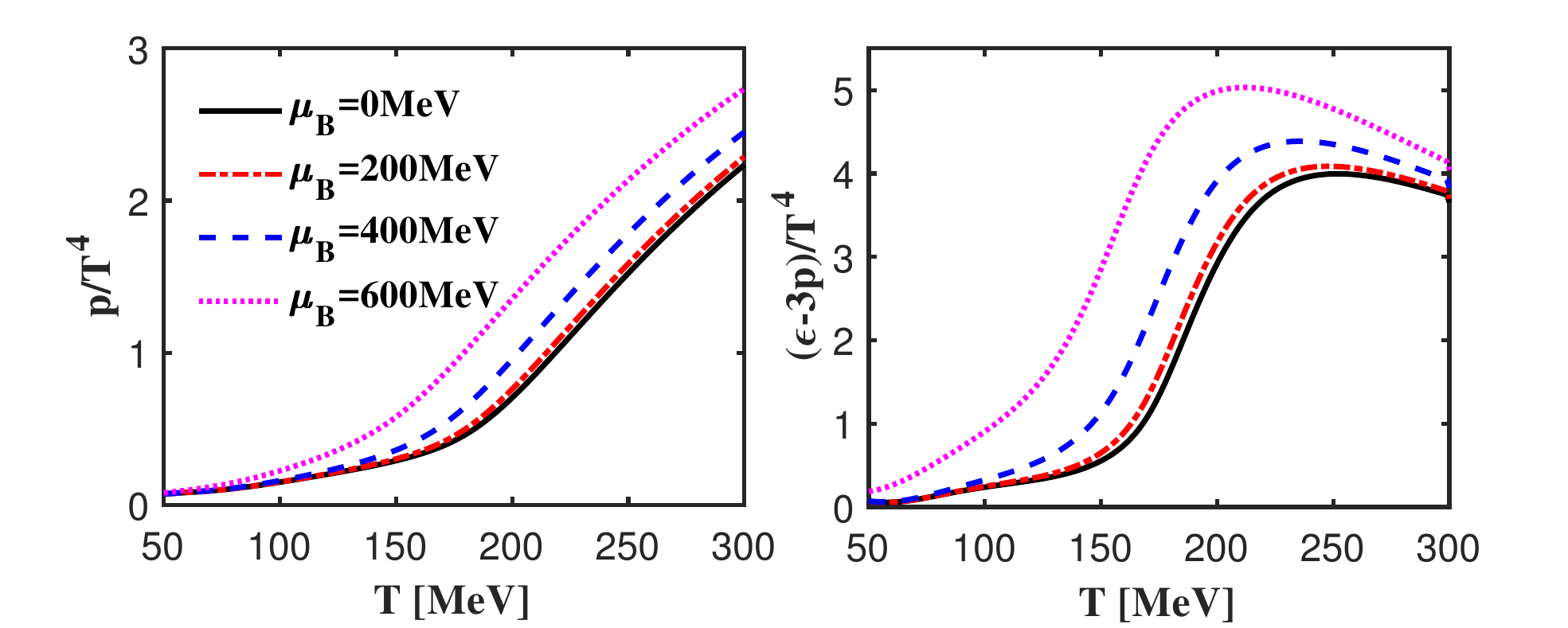}
\caption{Pressure (left panel) and trace anomaly (right panel) as functions of the temperature at several values of the baryon  chemical potential, where the Haar glue potential in \Eq{eq:GluepHaar} with $T_c^\text{\tiny{glue}}=270$ MeV and $\alpha=0.52$ is used.}\label{fig:mu1}
\end{figure*}
%

%
\begin{figure*}[t]
\includegraphics[width=1.\textwidth]{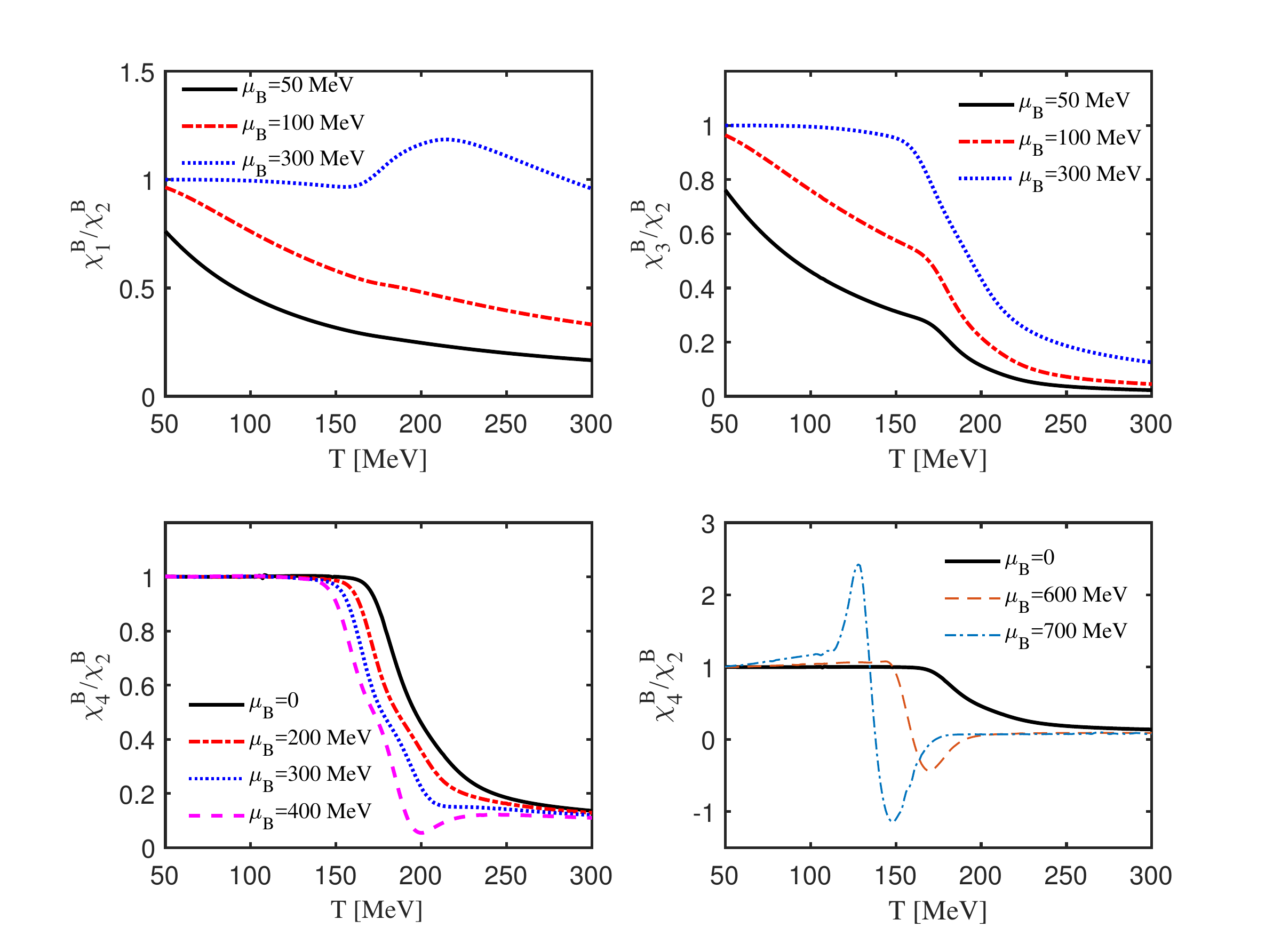}
\caption{Ratios of baryon number fluctuations: $\chi_1^{\text{B}}/\chi_2^{\text{B}}$ (top-left), $\chi_3^{\text{B}}/\chi_2^{\text{B}}$ (top-right), $\chi_4^{\text{B}}/\chi_2^{\text{B}}$ (bottom) as functions of the temperature at several values of the baryon chemical potential, where the Haar glue potential in \Eq{eq:GluepHaar} with $T_c^\text{\tiny{glue}}=270$ MeV and $\alpha=0.52$ is used.}\label{fig:mu2}
\end{figure*}
%

In this section we would like to give our numerical results, but before that, the initial conditions of the flow equation for the effective potential in \Eq{eq:dtUk} as well as other parameters, e.g. those in \Eq{eq:tildeU}, have to be specified. They are fixed by fitting hadronic observables at vacuum, and relevant discussions are presented in \app{app:num} in detail, and we also discuss how to reduce the influence of the UV cutoff $\Lambda$ on observables at large $T$ or $\mu$ therein.

In \Fig{fig:mass} we show the meson and quark masses as functions of the RG scale $k$ at vacuum, and as functions of the temperature. When we refer to results at finite temperature or chemical potential,  the RG scale $k=0$ is assumed, which is still applicable in the following. Comparing the left and right panels in \Fig{fig:mass}, one finds that the dependent behaviors of the masses on $k$ and the temperature are similar, which are reasonable and self-consistent.  With the increase of $k$ or $T$, the dynamically broken chiral symmetry is restored, and the chiral partners, such as $\pi$-$\sigma$, $a_0$-$\eta^{\prime}$, $K$-$\kappa$, become degenerate with each other, and furthermore, the constituent quark masses decrease pronouncedly, especially the light quarks. Note, however, that if we compare the results obtained here in the low energy effective model with those in the QCD calculations, including the quantum fluctuations of the glue sector, e.g. in \cite{Braun:2014ata,Fu:2018a}, the distinction is remarkable, and one would find that the mesons in the QCD calculations decouples much more quickly than the results here, once the scale or the temperature is located in the chiral symmetrical phase. And also the quark masses approach their bare masses more slowly in the effective model than those in the QCD. This feature of the effective model will affect the thermodynamics to be discussed in the following. 

In \Fig{fig:fpifK} we show the pion and kaon decay constants as functions of $T$ with $\mu_B=0$. As shown in \Eq{eq:fpiK}, $f_\pi$ and $f_K$ are linked to the expected values of the  $\sigma_L$ and $\sigma_S$ fields, and therefore serve as order parameters for the QCD chiral phase transition, or more exactly the continuous crossover. Alternatively, one can also use $\rho_1$ defined in \Eq{eq:rho1} as the order parameter, which is reduced to
\begin{align}
  \rho_1&=\frac{1}{2}(\bar{\sigma}_L^2+\bar{\sigma}_S^2)\,,\label{eq:rho1bar}
\end{align}
after expected values for all mesonic fields have been inserted, and note that only those of $\sigma$ fields are nonvanishing. The pseudocritical temperature extracted from the peak of $|\partial \rho_1/ \partial T|$ at $\mu_B=0$ is $T_c^{\chi}=194$ MeV, where the superscript $\chi$ denotes the chiral crossover, which is intended to be distinguished from the pseudocritical temperature for the deconfinement phase transition $T_c^{d}$. It is found in our calculations that $T_c^{d}=177$ MeV, where $T_c^{d}$ is obtained from the peak of the derivative of the Polyakov loop w.r.t. the temperature. Note that $T_c^{\chi}$ in this work is larger than that of lattice QCD by HotQCD Collaboration $154 \pm 9$ MeV  \cite{Bazavov:2014pvz} and Wuppertal-Budapest Collaboration $156 \pm 9$ MeV \cite{Borsanyi:2010bp,Borsanyi:2013bia}. We think this remarkable difference is attributed to several reasons as follow. Firstly, the absolute scale of the effective model is inherently different from that of QCD, and the former is larger. A larger critical temperature is also found in another effective model calculation with some different setup \cite{Fu:2018qsk}. Secondly, in the calculations we have used the glue potential $V_{\text{glue-Haar}}$ in \Eq{eq:GluepHaar} with two parameters $T_c^\text{\tiny{glue}}=270$ MeV and $\alpha=0.52$, and although the value of $T_c$ is acceptable, but a bit larger. Finally, the $\sigma$-meson mass is fixed to be $m_{\sigma}=510$ MeV in our calculations, as discussed in detail in \app{app:num}. This value is located in the experimental measured mass region of $f_0(500)$, i.e., $400-550$ MeV \cite{Patrignani:2016xqp}, but almost touches the allowed maximal value. However, because of the numerical instability for the Taylor expansion of the effective potential around the physical point in \Eq{eq:UkTaylor}, see \app{app:flowV} for more relevant discussions, it is difficult to decrease $m_{\sigma}$ further. A smaller $\sigma$-meson mass will results in a smaller $T_c^{\chi}$. We will try to overcome this numerical instability and present relevant results elsewhere.

Although there is an absolute scale difference between the effective model and the lattice simulation, it will not hamper the comparisons of results obtained from the two calculations, if the relative scale is used, such as the temperature in unit of the critical one, as will be discussed in what follows.

We show the pressure and the trace anomaly in \Fig{fig:PT4Tr} and the fluctuations of the baryon number up to the sixth order in \Fig{fig:R4262}. Two different glue potentials in \app{app:gluepot} are used in the calculations. Our calculated results are also compared with relevant lattice results by HotQCD Collaboration \cite{Bazavov:2014pvz,Bazavov:2017dus,Bazavov:2017tot} and Wuppertal-Budapest Collaboration \cite{Borsanyi:2013bia,Borsanyi:2013hza}. In order to eliminate the difference of absolute scale between the lattice QCD and the effective model as discussed above, we rescale the temperature by their respective pseudocritical temperature. The pseudocritical temperature $T_c$ for the lattice calculations is chosen to be central value, i.e., 154 MeV from $154 \pm 9$ MeV by HotQCD Collaboration  \cite{Bazavov:2014pvz}, and 156 MeV from $156 \pm 9$ MeV by Wuppertal-Budapest Collaboration \cite{Borsanyi:2010bp,Borsanyi:2013bia}, and the errors are not taken into account in this work. Note that, in the low energy effective model, two pseudocritical temperatures $T_c^{\chi}$ and $T_c^{d}$ are defined, which are related to the chiral and deconfinement phase transitions, respectively. In \Fig{fig:PT4Tr} we choose $T_c=T_c^{\chi}$ for the effective model. One finds that our calculated pressure and trace anomaly agree with the lattice results. This is not a surprise, since two parameters in the glue potential are employed to fit the pressure and the trace anomaly, see \app{app:gluepot} for more detailed discussions. Therefore, it is more valuable to compare our calculated baryon number fluctuations with the lattice simulations as shown in \Fig{fig:R4262}. However, there is still an intricacy needed to be fixed before the comparison. It has been known that the ratio $\chi_4^{\text{B}}/\chi_2^{\text{B}}$ is linked to the degrees of freedom \cite{Ejiri:2005wq,Fu:2009wy,Skokov:2010wb,Skokov:2010uh,Karsch:2011gg,Fu:2015naa}, and therefore this ratio is more sensitive to the deconfinement phase transition. Thus in \Fig{fig:R4262} we relax $T_c$ of the effective model to be an appropriate value between $T_c^{\chi}$ and $T_c^{d}$, and we find $T_c=185$ MeV for the Haar potential which gives an optimal agreement. This is nontrivial, and it would be clearer if one looks at the red dashed curve calculated with the polynomial glue potential. The best choice for the polynomial potential is $T_c=T_c^{\chi}$. Nevertheless, the agreement with the lattice results for the polynomial potential is not as good as that for the Haar potential, specifically at high temperature. The ratio between the sixth and the quadratic fluctuations of the net baryon number is also presented in the right panel of \Fig{fig:R4262}, in comparison to the lattice results. One finds that the prediction of the low energy effective model is in qualitative agreement with the lattice simulation within the considerable errors.

In \Fig{fig:R4262usK}, we show the fluctuations of the light and strange quarks as functions of the temperature, where $\chi_4/\chi_2$ and $\chi_6/\chi_2$ are presented in the left and right panels, respectively. Apparently, the magnitude of the fluctuations for light quarks is larger than that for the strange quark, since the strange quark has larger mass. Moreover, one observes that the ratio between the sixth and the second order fluctuations for the strange quark, i.e., $\chi_6^s/\chi_2^s$ denoted by the red solid line in the right panel of \Fig{fig:R4262usK}, develops a rising behavior when the temperature is above the critical temperature. In order to find the reason behind, we turn off the quantum fluctuations resulting from the $K$ and $\kappa$ loops in \Eq{eq:dtUk}, and the relevant results are shown in \Fig{fig:R4262usK} with the dash-dotted lines. After the quantum fluctuations of $K$ and $\kappa$ mesons are removed, $\chi_6^s/\chi_2^s$ drops pronouncedly in the high temperature regime. Note that, in the low energy effective model, the masses of mesons grow slowly with $T$ when the temperature is above the critical temperature, as shown in \Fig{fig:mass}, and that is the reason why degrees of freedom of the $K$ and $\kappa$ mesons still play a role in $\chi_6^s/\chi_2^s$ when the temperature is relatively high. However, we should mention that recent studies from QCD indicate that, unlike the effective model, the mesons decouple from the system quickly once the temperature is above $T_c$ \cite{Fu:2019b}, which will relieve the problem of $\chi_6^s/\chi_2^s$ at high temperature. Relevant studies will be done in the future. Furthermore, we also find that quantum fluctuations of $K$ and $\kappa$ mesons decrease the $\chi_4/\chi_2$ and $\chi_6/\chi_2$ during the crossover, i.e. in the region of 150 MeV $\lesssim T\lesssim$ 200 MeV, both for light and strange quarks.

In this work, we also perform calculations at finite baryon chemical potential. The phase diagram of the 2+1 flavor low energy effective model in the plane of $T$ and $\mu_B$ is shown in \Fig{fig:drhodT}. We use the color to indicate the value of $\big\vert\partial \rho_1(T,\mu_B)/\partial T \big\vert$ with the chiral order parameter $\rho_1$ given in \Eq{eq:rho1bar}. The two solid lines are determined by 90\% peak height of this derivative at each value of $\mu_B$. This two lines approach toward each other with the increase of the baryon chemical potential, and crosses at the critical end point. The location of the CEP is found to be ($T_{\text{\tiny{CEP}}}=84\,\text{MeV}\,,{\mu_B}_{\text{\tiny{CEP}}}=840\,\text{MeV}$). The method of Taylor expansion of the effective potential results in numerical instability around the first-order phase transition, so we do not show results at very high baryon chemical potential. The ideal approach to reveal the global properties of the effective potential is to lattice the potential in the field space, which therefore can be employed to study the first-order phase transition. Interested readers are referred to, e.g., \cite{Herbst:2013ail}.

In \Fig{fig:mu1} we investigate the dependence of the pressure and trace anomaly on the baryon chemical potential. One finds that the pressure scaled by $T^4$ as a function of the temperature moves up globally with the increase of $\mu_B$, and the height of the trace anomaly increases as well. Note that our calculated results are obtained with a glue potential without direct dependence on the baryon chemical potential. When the influence of the chemical potential on the glue potential is taken into account, for instance, through the $\mu$-modification of the parameter $T_c^\text{\tiny{glue}}$ in the glue potential \cite{Fu:2018qsk}, It is reasonable to expect that the behavior of the trace anomaly would change considerably, cf. \Fig{fig:Tcgluo}. Unfortunately, a conclusion about the dependence of the glue potential on the chemical potential has not yet been arrived at, but it might be inferred through a detailed comparison of the trace anomaly at finite $\mu_B$ between the low energy effective theory and the lattice simulations, which is though very interesting but beyond the scope of this work, and we will report it elsewhere in the future.

In \Fig{fig:mu2} we show $\chi_1^{\text{B}}/\chi_2^{\text{B}}$, $\chi_3^{\text{B}}/\chi_2^{\text{B}}$, $\chi_4^{\text{B}}/\chi_2^{\text{B}}$ as functions of the temperature at several values of the baryon chemical potential. $\chi_1^{\text{B}}/\chi_2^{\text{B}}$ and $\chi_3^{\text{B}}/\chi_2^{\text{B}}$ are linearly dependent on $\mu_B$ when $\mu_B$ is not far from zero, so they are sensitive to the chemical potential in this region, which is also verified in our calculations. Therefore, $\chi_1^{\text{B}}/\chi_2^{\text{B}}$ or $\chi_3^{\text{B}}/\chi_2^{\text{B}}$ is usually employed to extract the freeze-out chemical potential \cite{Borsanyi:2013hza,Bazavov:2017tot}. Furthermore, one finds that with the increase of the baryon chemical potential, the kurtosis of the baryon number distribution, i.e., the ratio $\chi_4^{\text{B}}/\chi_2^{\text{B}}$ develops a minus value as well as a peak, which is more significant when the critical end point is approached. But the region of minus $\chi_4^{\text{B}}/\chi_2^{\text{B}}$ shrinks and vanishes at the CEP \cite{Schaefer:2011ex}.


\section{Summary and outlook}
\label{sec:sum}

We have studied the QCD phase transition, thermodynamics, and fluctuations of the baryon number and strangeness in the 2+1 flavor low energy effective model. In the calculations quantum fluctuations are included through the functional renormalization group approach. The flow equation for the effective potential is solved by Taylor-expanding it around the physical point.

The QCD phase transition, as well as the behaviors of the masses of mesons and quarks, the pion and kaon decay constants during the phase transition, has been investigated in the 2+1 flavor effective model at finite temperature and chemical potential.

The equation of state for the QCD matter, including the pressure and the trace anomaly, and the fluctuations of the baryon number  up to the sixth order are calculated and compared with lattice results. We find the agreement between the low energy effective model and the lattice QCD is acceptable, especially for the calculations with the glue potential which takes the Polyakov loop fluctuations into account. The fluctuations of light and strange quarks are also computed up to the sixth order, and the fluctuation of the strange quark is less than that of the light quarks. We also calculate the EoS and baryon number fluctuations at finite baryon chemical potential.

It should be noted that this work is our first calculation of high order cumulants of the conserved charge distribution in the 2+1 flavor low energy effective model within the FRG approach. Lots of things will have to be done in the future, for instance, going beyond the LPA truncation employed in this work; including the degree of freedom of gluons and extending the low energy effective theory to the 2+1 flavor rebosonized QCD. Furthermore, in our calculations the constraints of the strange neutrality and a fixed $Z/A$ are not implemented,  where $Z$ and $A$ are electric and mass number for a nucleus. These constraints are quite relevant to the experiments, see, e.g. \cite{Fu:2018qsk} for details, so they should also be taken into account in our studies in the future.

\begin{acknowledgments}

We thank Jan M. Pawlowski and Fabian Rennecke for valuable discussions, and Heng-Tong Ding for providing us with lattice data. We are grateful to Bernd-Jochen Schaefer and Heng-Tong Ding for reading this manuscript and giving us a number of valuable comments and suggestions. We also thank the members of the fQCD collaboration \footnote{{\it{fQCD Collaboration}}, J. Braun, L. Corell, A. K. Cyrol,
W.-j. Fu, C. Huang, M. Leonhardt, M. Mitter, J. M. Pawlowski, M. Pospiech, F. Rennecke, C. Schneider, R. Wen, N. Wink, S. Yin.} for work on related projects. The work was supported by the National Natural Science Foundation of China under Contracts Nos. 11775041.

\end{acknowledgments}


\appendix

\section{Flow of the effective potential}
\label{app:flowV}

%
\begin{figure*}[t]
\includegraphics[width=1.\textwidth]{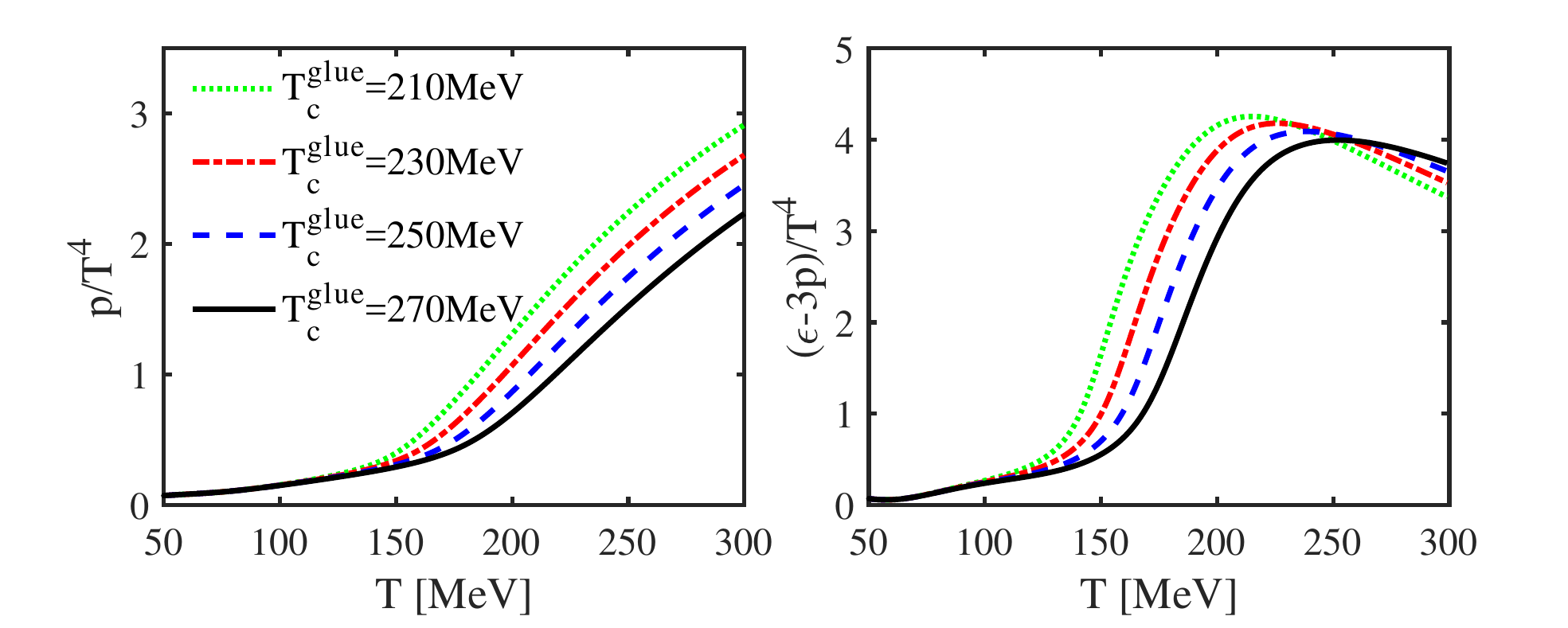}
\caption{Pressure (left panel) and trace anomaly (right panel) as functions of the temperature at vanishing chemical potential, where we have compared the results with different values of $T_c^\text{\tiny{glue}}$ in the glue potential $V_{\text{glue-Haar}}$ in \Eq{eq:GluepHaar}, and $\alpha=0.52$ is fixed.}\label{fig:Tcgluo}
\end{figure*}
%

%
\begin{figure*}[t]
\includegraphics[width=1.\textwidth]{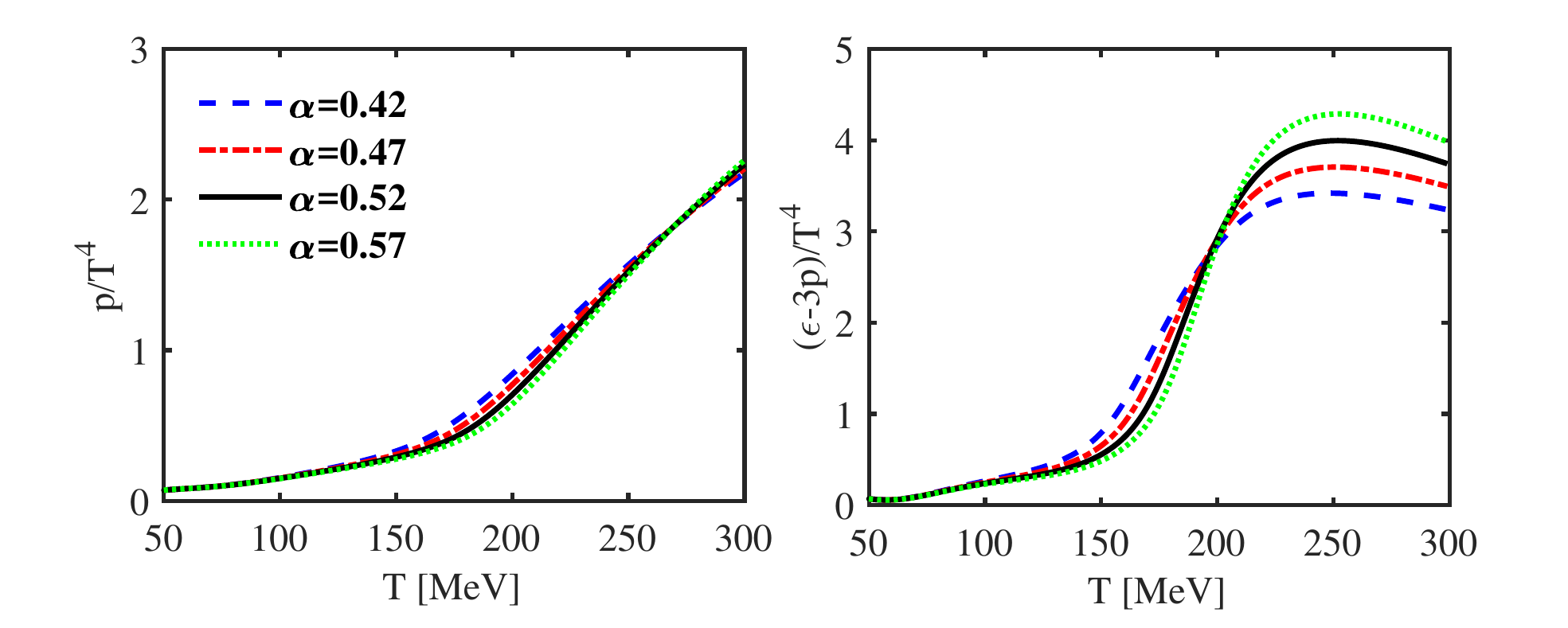}
\caption{Same as \Fig{fig:Tcgluo} but with $T_c^\text{\tiny{glue}}=270$ MeV fixed and several values of $\alpha$.}\label{fig:alpha}
\end{figure*}
%

The flow equation of the effective potential in \Eq{eq:dtUk} is obtained with the flat or Litim $3d$ IR regulators \cite{Litim:2000ci,Litim:2001up}, as follow
\begin{align}
  R^{q}_{k}(q_{0},\bm{q})&=i\bm{\gamma}\cdot\bm{q}\,r_{F}(\bm{q}^{2}/k^2)\,,\\[2ex] 
  R^{\phi}_{k}(q_{0},\bm{q})&=\bm{q}^{2}r_{B}(\bm{q}^{2}/k^2)\,, \label{}
\end{align}
for the quark and meson fields, respectively, where the shape functions read
\begin{align}
  r_{F}(x)&=\Big(\frac{1}{\sqrt{x}}-1\Big)\Theta(1-x)\,,\\[2ex] 
  r_{B}(x)&=\Big(\frac{1}{x}-1\Big)\Theta(1-x)\,,  \label{}
\end{align}
with the Heaviside step function $\Theta(x)$. 

The threshold functions in \Eq{eq:dtUk} are given by
\begin{align}
  l_{0}^{(B)}(m^2;T,\mu)&=\frac{1}{3\sqrt{1+m^{2}}}\Big(1+n_{B}(m^{2};T,\mu)\nonumber\\[2ex]
	&\hspace{.3cm}+n_{B}(m^{2};T,-\mu)\Big)\,, \label{eq:l0B}\\[2ex]
  l_{0}^{(F)}(m^2;T,\mu)&=\frac{1}{3\sqrt{1+m^{2}}}\Big(1-n_{F}(m^{2};T,\mu,L,\bar L)\nonumber\\[2ex]
    &\hspace{.3cm}-n_{F}(m^{2};T,-\mu,\bar L,L)\Big)\,,\label{eq:l0F}
\end{align}
where the bosonic distribution function reads
\begin{align}
	n_{B}(m^{2};T,\mu)=&\frac{1}{e^{(k\sqrt{1+m^{2}}-\mu)/T}-1}\,,
\end{align}
and the Polyakov-loop modified fermionic distribution function
\begin{align}
  n_{F}(m^{2};T,\mu,L,\bar L)=&\frac{1+2\bar L\,e^{x/T}+L\,e^{2x/T}}{1+3\bar L\,e^{x/T}+3L\, e^{2x/T}+e^{3x/T}}\,, \label{eq:nFL}
\end{align}
with 
\begin{align}
  x=&k\sqrt{1+m^{2}}-\mu\,. \label{}
\end{align}

In this work, we employ the method of Taylor expansion to solve the flow equation for the effective potential in \Eq{eq:dtUk}, which reads
\begin{align}
  U_k(\rho_1,\tilde{\rho}_2)=&\sum_{m,n=0}^{N}\frac{\lambda_{mn,k}}{m!\,n!}(\rho_1-\kappa_{1,k})^m(\tilde{\rho}_2-\kappa_{2,k})^n\,, \label{eq:UkTaylor}
\end{align}
with $N\geq m+2n$ being the maximal order of the expansion, and $N=5$ is adopted in this work, which is found to suffice for the convergence of calculations. $\kappa_{1,k}$ and $\kappa_{2,k}$ are the expansion points for $\rho_1$ and $\tilde{\rho}_2$, respectively. The expansion points usually can be chosen with a freedom given being in the convergence regime, in which the physical results are independent of the choice of the expansion points. In literatures there are two commonly employed choices. One is the fixed point expansion with the expansion points, i.e., $\kappa_{1,k}$ and $\kappa_{2,k}$ here,  independent of the RG scale $k$ \cite{Pawlowski:2014zaa,Rennecke:2016tkm,Fu:2018qsk}. The prominent advantage of the fixed point expansion lies in its excellent numerical stability. The other is the physical point expansion, see, e.g. \cite{Sun:2018ozp}, in which $\kappa_{1,k}$ and $\kappa_{2,k}$ are the physical points for every value of $k$, and therefore they are $k$ dependent. As the convergence is concerned, the physical point expansion is superior to the fixed point expansion, while the former loses the excellent numerical stability.

In this work we employ the approach of physical point expansion. Differentiating both sides of \Eq{eq:UkTaylor} w.r.t. the RG time $t$, one arrives at
\begin{align}
  \partial_t U_k(\rho_1,\tilde{\rho}_2)&=\sum_{m,n=0}^{N}\frac{1}{m!\,n!}\Big(\partial_t\lambda_{mn,k}-\lambda_{m+1\,n,k}\partial_t \kappa_{1,k}\nonumber\\[2ex]
&-\lambda_{m\,n+1,k}\partial_t \kappa_{2,k}\Big)(\rho_1-\kappa_{1,k})^m(\tilde{\rho}_2-\kappa_{2,k})^n, \label{}
\end{align}
which yields
\begin{align}
  &\partial_{\rho_1}^{m}\partial_{\tilde{\rho}_2}^{n}\Big(\partial_t U_k(\rho_1,\tilde{\rho}_2)\Big)\bigg\vert_{\substack{\rho_1=\kappa_{1,k}\\ \tilde{\rho}_2=\kappa_{2,k}}}\nonumber\\[2ex]
  =&\partial_t\lambda_{mn,k}-\lambda_{m+1\,n,k}\partial_t \kappa_{1,k}-\lambda_{m\,n+1,k}\partial_t \kappa_{2,k}\,. \label{}
\end{align}
Specially, for the expansion coefficients $\lambda_{10,k}$ and $\lambda_{01,k}$, one has
\begin{align}
  &\partial_t\lambda_{10,k}-\lambda_{20,k}\partial_t \kappa_{1,k}-\lambda_{11,k}\partial_t \kappa_{2,k}\nonumber\\[2ex]
  &-\partial_{\rho_1}\big(\partial_t U_k(\rho_1,\tilde{\rho}_2)\big)\Big\vert_{\substack{\rho_1=\kappa_{1,k}\\ \tilde{\rho}_2=\kappa_{2,k}}}=0\,,\label{eq:dtlam10}\\[2ex]
  &\partial_t\lambda_{01,k}-\lambda_{11,k}\partial_t \kappa_{1,k}-\lambda_{02,k}\partial_t \kappa_{2,k}\nonumber\\[2ex]
  &-\partial_{\tilde{\rho}_2}\big(\partial_t U_k(\rho_1,\tilde{\rho}_2)\big)\Big\vert_{\substack{\rho_1=\kappa_{1,k}\\ \tilde{\rho}_2=\kappa_{2,k}}}=0\,.\label{eq:dtlam01}
\end{align}
In the meantime, in order to find the evolution equations for the expansion points, the stationarity condition is implemented as follows
\begin{align}
  \frac{\partial \tilde{U}_{k}(\sigma_L,\sigma_S)}{\partial \sigma_L}\bigg\vert_{\substack{\sigma_L=\bar \sigma_L\\ \sigma_S=\bar \sigma_S}}=0,\quad  \frac{\partial \tilde{U}_{k}(\sigma_L,\sigma_S)}{\partial \sigma_S}\bigg\vert_{\substack{\sigma_L=\bar \sigma_L\\ \sigma_S=\bar \sigma_S}}=0,\label{}
\end{align}
which leads us to the following equations which read
\begin{align}
  &\big[\lambda_{10,k}+\frac{1}{6}(3\bar{\sigma}_L^2-2\bar{\sigma}_S^2)\lambda_{01,k}-\frac{1}{\sqrt{2}}c_A\bar{\sigma}_S\big]\partial_t \bar{\sigma}_L\nonumber\\[2ex]
  &+\big[-\frac{2}{3}\bar{\sigma}_L \bar{\sigma}_S \lambda_{01,k}-\frac{1}{\sqrt{2}}c_A\bar{\sigma}_L\big]\partial_t \bar{\sigma}_S+\bar{\sigma}_L \partial_t \lambda_{10,k}\nonumber\\[2ex]
  &+\frac{\bar{\sigma}_L}{6}(\bar{\sigma}_L^2-2\bar{\sigma}_S^2) \partial_t \lambda_{01,k}=0\,,\label{eq:dtdsigmalUk}
\end{align}
and
\begin{align}
  &\big[-\frac{2}{3}\bar{\sigma}_L \bar{\sigma}_S \lambda_{01,k}-\frac{1}{\sqrt{2}}c_A\bar{\sigma}_L\big]\partial_t \bar{\sigma}_L+\big[\lambda_{10,k}\nonumber\\[2ex]
  &-\frac{1}{3}(\bar{\sigma}_L^2-6\bar{\sigma}_S^2)\lambda_{01,k}\big]\partial_t \bar{\sigma}_S+\bar{\sigma}_S\partial_t \lambda_{10,k}\nonumber\\[2ex]
  &-\frac{\bar{\sigma}_S}{3}(\bar{\sigma}_L^2-2\bar{\sigma}_S^2) \partial_t \lambda_{01,k}=0\,.\label{eq:dtdsigmasUk}\end{align}
It follows from \Eq{eq:rho1} and \Eq{eq:rho2} that one arrives at
\begin{align}
  \kappa_{1,k}&=\frac{1}{2}(\bar{\sigma}_L^2+\bar{\sigma}_S^2)\,,\label{eq:kappa1}\\[2ex]
  \kappa_{2,k}&=\frac{1}{24}(\bar{\sigma}_L^2-2\bar{\sigma}_S^2)^2\,,\label{eq:kappa2}
\end{align}
which result in
\begin{align}
  \partial_t \kappa_{1,k}&=\bar{\sigma}_L\partial_t \bar{\sigma}_L+\bar{\sigma}_S\partial_t \bar{\sigma}_S\,,\label{eq:dtkappa1}\\[2ex]
  \partial_t \kappa_{2,k}&=\frac{1}{6}\big(\bar{\sigma}_L^2-2\bar{\sigma}_S^2\big)\big(\bar{\sigma}_L\partial_t \bar{\sigma}_L-2\bar{\sigma}_S\partial_t \bar{\sigma}_S\big)\,.\label{eq:dtkappa2}
\end{align}
Equations (\ref{eq:dtlam10}),  (\ref{eq:dtlam01}),  (\ref{eq:dtdsigmalUk}),  (\ref{eq:dtdsigmasUk}), (\ref{eq:dtkappa1}), (\ref{eq:dtkappa2}) constitute a closed set of linear equations, which can be solved straightforwardly to obtain the flows equations for $\kappa_{1,k}$, $\kappa_{2,k}$, $\lambda_{10,k}$, and $\lambda_{01,k}$. Since the final expressions for these flows are lengthy and the set of linear equations can be quite easily solved, we will not present them here.


\section{Glue potential}
\label{app:gluepot}

%
\begin{figure*}[t]
\includegraphics[width=1.\textwidth]{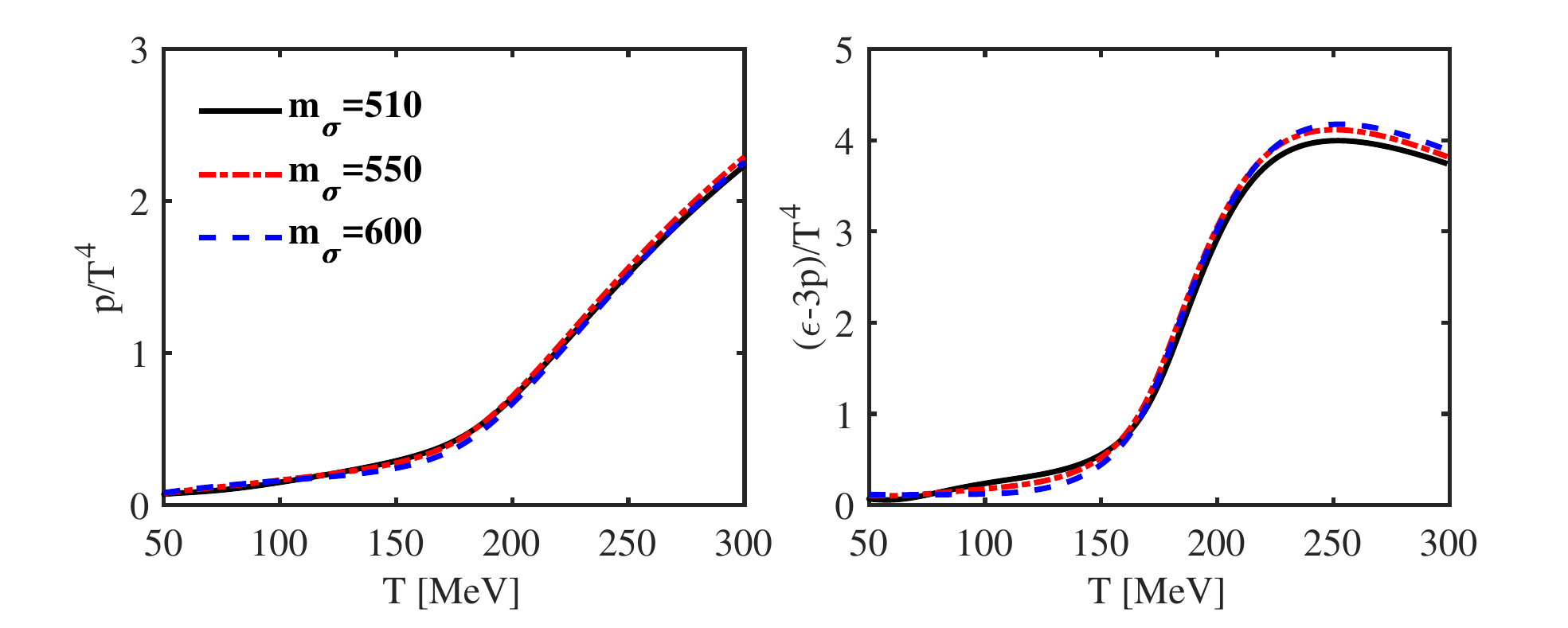}
\caption{Pressure (left panel) and trace anomaly (right panel) as functions of the temperature at $\mu_B=0$. We have compared calculations with different sets of parameters in the hadronic sector, which produce different values of the $\sigma$-meson mass at vacuum, while other hadronic observables are the same. The glue potential $V_{\text{glue-Haar}}$ with $T_c^\text{\tiny{glue}}=270$ MeV and $\alpha=0.52$ is used.}\label{fig:mSigma1}
\end{figure*}
%

%
\begin{figure*}[t]
\includegraphics[width=1.\textwidth]{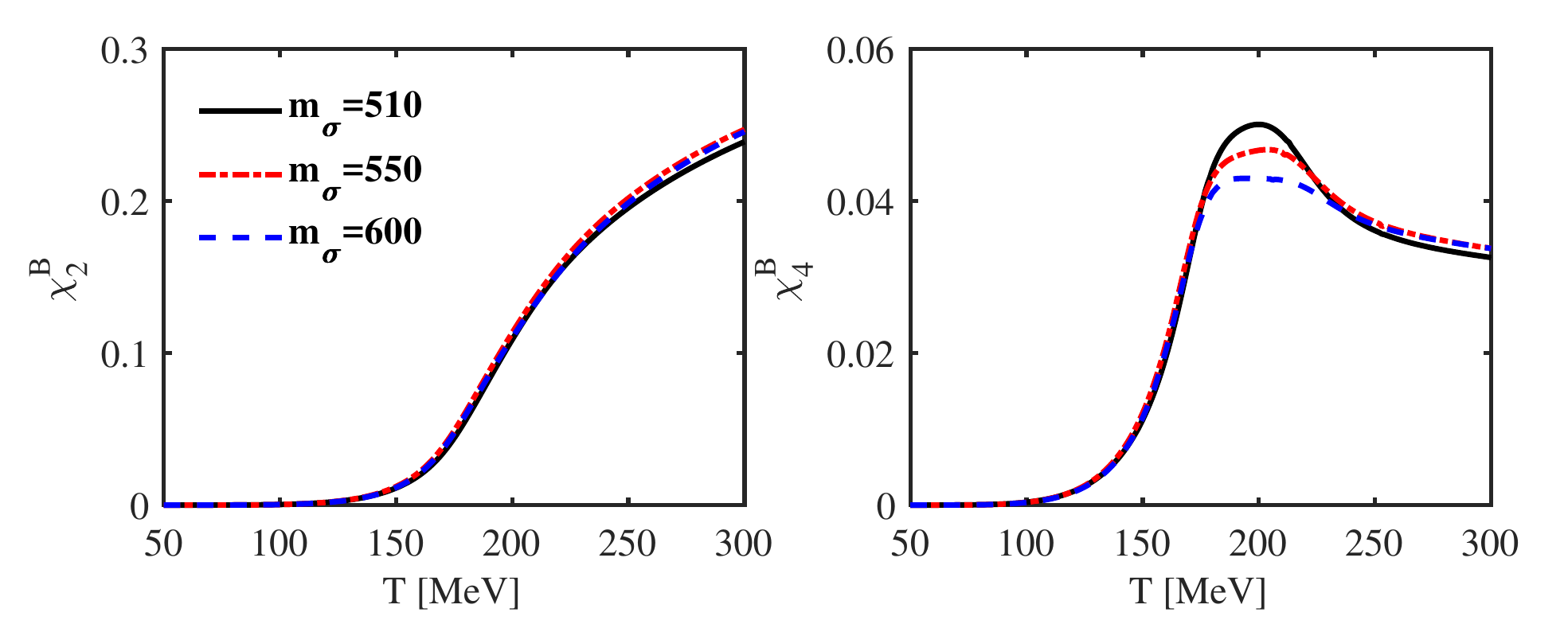}
\caption{Same as \Fig{fig:mSigma1} but for the quadratic (left panel) and quartic (right panel) baryon number fluctuations.}\label{fig:mSigma2}
\end{figure*}
%

%
\begin{figure}[t]
\includegraphics[width=0.5\textwidth]{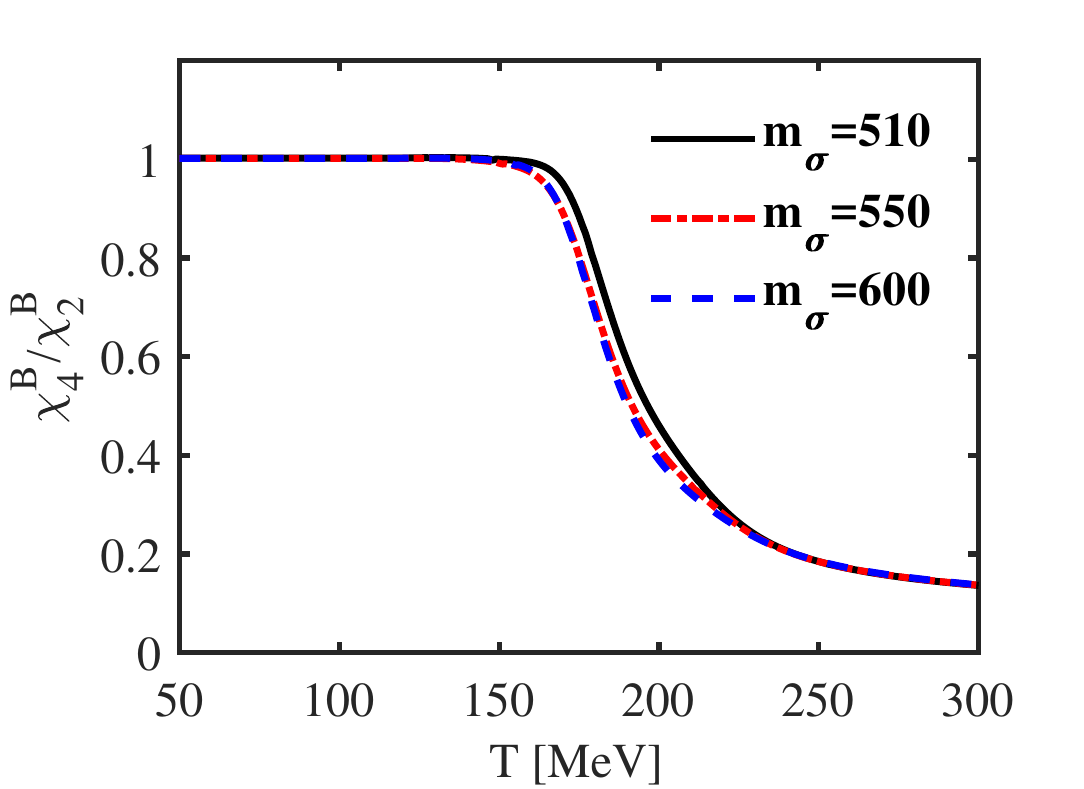}
\caption{Same as \Fig{fig:mSigma1} but for the kurtosis of the baryon number distribution, i.e., the ratio $\chi_4^B/\chi_2^B$.}\label{fig:mSigma3}
\end{figure}
%

In this work two different parameterizations for the glue potential, i.e., the Polyakov-loop potential $V_{\text{\tiny{glue}}}(L,\bar L)$  in \Eq{eq:action}, are employed. One is the polynomial potential, which is widely used in literatures, see e.g. \cite{Ratti:2005jh}, and given by
\begin{align}
  \bar V_{\text{{glue-poly}}}(L,\bar L)=& -\frac{b_2(T)}{2}L\bar{L}-\frac{b_3}{6}(L^3+\bar{L}^3)+\frac{b_4}{4}(L\bar{L})^2\,.\label{eq:Glueppoly}
\end{align}
with $ \bar V_{\text{\tiny{glue-poly}}}=V_{\text{\tiny{glue}}}/T^4$, and the temperature dependence is only encoded in the coefficient $b_2$ which reads
\begin{align}
  b_2(T)&=a_1+\frac{a_2}{1+t}+\frac{a_3}{(1+t)^2}+\frac{a_4}{(1+t)^3}\,,\label{eq:b2}
\end{align}
where $t\equiv (T-T_c)/T_c$ is the reduced temperature. Note that the Polyakov-loop potential is parameterized by employing the thermodynamics and the behaviors of Polyakov loop in the Yang-Mills (YM) theory, thus $T_c$ is the critical temperature for the deconfinement phase transition in the YM theory. Interestingly, it is found that the YM Polyakov-loop potential is also applicable in QCD \cite{Haas:2013qwp}, and the unquenching effect is quantitatively taken into account, given the reduced temperature is appropriately rescaled, for instance
\begin{align}
  t_{\text{\tiny{YM}}}&\rightarrow \alpha\,t_{\text{\tiny{glue}}}\,,\label{}
\end{align}
with
\begin{align}
  t_{\text{\tiny{glue}}}&=(T-T_c^\text{\tiny{glue}})/T_c^\text{\tiny{glue}}\,,\label{}
\end{align}
where $\alpha$ is the linear scale factor and $T_c^\text{\tiny{glue}}$ is the pseudocritical temperature for the deconfinement phase transition in QCD. It is found that $\alpha \simeq 0.57$ in the case of flavor $N_f=2$ \cite{Haas:2013qwp}. $T_c^\text{\tiny{glue}}$ can also be estimated by the dependence of $\Lambda_{\text{QCD}}$ on $N_f$ via the QCD RG running \cite{Schaefer:2007pw}, and $T_c^\text{\tiny{glue}}=$ 208 MeV for $N_f=2$ and 187 MeV for $N_f=2+1$. In this work we will relax the restriction on the values of parameters $\alpha$ and $T_c^\text{\tiny{glue}}$, and study the dependence of the QCD thermodynamics on them systematically.

In recent year, another parameterization for the Polyakov-loop potential is proposed \cite{Lo:2013hla}, in which fluctuations of the Polyakov loop are also taken into account besides usually employed quantities. Furthermore, this parameterization employs the  $SU(N_c)$ Haar measure, which solves the problem, as the polynomial potential in \Eq{eq:Glueppoly} has, that the Polyakov loop exceeds unity at high temperature. We denote this potential as $\bar V_{\text{glue-Haar}}$ which reads
\begin{align}
  \bar V_{\text{glue-Haar}} &= -\frac{\bar a(T)}{2} \bar L L + \bar b(T)\ln M_H(L,\bar{L})\nonumber \\[2ex]
  &\quad + \frac{\bar c(T)}{2} (L^3+\bar L^3) + \bar d(T) (\bar{L} L)^2\,,\label{eq:GluepHaar}
\end{align}
with the Haar measure 
\begin{align}
M_H (L, \bar{L})&= 1 -6 \bar{L}L + 4 (L^3+\bar{L}^3) - 3  (\bar{L}L)^2\,.
\end{align}
where coefficients in \Eq{eq:GluepHaar} are hatted with a bar in order to be distinguished from those in \Eq{eq:Glueppoly}. Different from the polynomial potential, all the coefficients in \Eq{eq:GluepHaar} are dependent on the temperature, and their dependence is given by
\begin{align}
  x(T) &= \frac{x_1 + x_2/(t+1) + x_3/(t+1)^2}{1 + x_4/(t+1) + x_5/(t+1)^2}\,,\label{eq:xT}
\end{align}
for $x\in \{\bar a, \bar c, \bar d\}$, and 
\begin{align}
  \bar b(T) &=\bar b_1 (t+1)^{-\bar b_4}\left (1 -e^{\bar b_2/(t+1)^{\bar b_3}} \right)\,.\label{eq:bT}
\end{align}
Values of all these coefficients as well as those in the Eqs. (\ref{eq:Glueppoly}) and (\ref{eq:b2}) are collected in \Tab{tab:coeffs}.

%
\begin{table}[tb!]
  \centering
  \begin{tabular}{cccccc}
    \hline\hline
    & 1 & 2 & 3 & 4 & 5 \rule{0pt}{2.6ex}\rule[-1.2ex]{0pt}{0pt}
    \\ \hline
    $\bar a_i$ &-44.14& 151.4 & -90.0677 &2.77173 &3.56403 \\
    $\bar b_i$ &-0.32665 &-82.9823 &3.0 &5.85559  &\\
    $\bar c_i$ &-50.7961 &114.038 &-89.4596 &3.08718 &6.72812\\
    $\bar d_i$ &27.0885 &-56.0859 &71.2225 &2.9715 &6.61433\\\hline
    $a_i$  &6.75 &-1.95 &2.625 &-7.44 &          \\
    $b_i$  &        &         &0.75  &7.5  &  \\\hline\hline
  \end{tabular}
  \caption{Constants in Eqs. (\ref{eq:Glueppoly}), (\ref{eq:b2}), (\ref{eq:xT}), (\ref{eq:bT}) for two different parameterizations of the glue potential.} 
  \label{tab:coeffs}
\end{table}
%

In \Fig{fig:Tcgluo} and \Fig{fig:alpha} we have investigated the dependence of the thermodynamics on the two parameters  $T_c^\text{\tiny{glue}}$ and $\alpha$ in the glue potential, respectively. One finds that $T_c^\text{\tiny{glue}}$ shifts the curves of the dimensionless pressure and trace anomaly along the $T$ direction, and affects the height of the trace anomaly as well, while variation of $\alpha$ only result in the change of height of the trace anomaly. Therefore, we would like to employ these properties to determine $\alpha$ and $T_c^\text{\tiny{glue}}$, guided by the lattice results of the pressure and the trace anomaly. This procedure would certainly reduce the predictive power of the effective model, when the pressure and trace anomaly with $\mu_B=0$ are concerned, but it does not, if we discuss other quantities, such as the baryon number fluctuations, calculations at finite chemical potentials, etc. In line with this idea, $T_c^\text{\tiny{glue}}=250$ MeV and $\alpha=0.54$ are determined for the polynomial potential in \Eq{eq:Glueppoly}, and $T_c^\text{\tiny{glue}}=270$ MeV and $\alpha=0.52$ for the Haar potential in \Eq{eq:GluepHaar}.


\section{Numerical setup}
\label{app:num}

In an ideal case, we wish to start the evolution of the FRG flows at an initial scale $\Lambda$ well above the scale of interests, such as $\sim2\pi T$ related to the Matsubara gap of the temperature. In a low energy effective theory this, however, is restricted by the lack of the glue dynamics as we discuss in \sec{sec:FRG}, since the gluon mass gap disappears rapidly when the RG scale is above $\sim$ 1 GeV \cite{Mitter:2014wpa,Braun:2014ata,Cyrol:2016tym,Cyrol:2017ewj}. Therefore, the value of $\Lambda$ is limited to a small window around $\sim$ 1 GeV. In this work without loss of the generality, we choose $\Lambda=1$ GeV.

At the initial scale $\Lambda$, the effective potential in \Eq{eq:dtUk} is classical, and it is reasonable to assume that its irrelevant terms, i.e. those with couplings of dimension minus,  in the Taylor expansion in \Eq{eq:UkTaylor} are vanishing, which leads us to
\begin{align}
  U_{k=\Lambda}(\rho_1,\tilde{\rho}_2)&=\lambda_{10,\Lambda}\rho_1+\frac{\lambda_{20,\Lambda}}{2}\rho_1^2+\lambda_{01,\Lambda}\tilde{\rho}_2\,,\label{}
\end{align}
where the three coefficients together with those in \Eq{eq:tildeU} and the Yukawa coupling constitute the set of parameters in the hadronic sector in the low energy effective theory. Their values used in this work are collected in \Tab{tab:paras}. With these values, one obtains hadronic observables in what follows: the pion and kaon decay constants and masses, $f_\pi=92$ MeV, $f_K=115$ MeV, $m_\pi=135$ MeV, $m_K=492$ MeV, the $\sigma$-meson mass, $m_\sigma=510$ MeV, the sum of mass square of $\eta$ and $\eta^\prime$ mesons, $m_{\eta}^2+m_{\eta^\prime}^2=1.22\,\text{GeV}^2$, the dressing masses of light and strange quarks $m_l=297$ MeV, $m_s=453$ MeV.

%
\begin{table}[tb!]
  \centering
  \begin{tabular}{ccccccc}
    \hline\hline
    $\lambda_{10,\Lambda} [\text{GeV}^2]$& $\lambda_{20,\Lambda}$ & $\lambda_{01,\Lambda}$ & $c_A [\text{GeV}]$ & $j_L [\text{GeV}^3]$ & $j_S [\text{GeV}^3]$ & h
    \\ \hline
    $(0.654)^2$  &   31    &  11.9  & 4.808 & $(0.119)^3$  & $(0.337)^3$ & 6.5 \\\hline\hline
  \end{tabular}
  \caption{Initial conditions of the flow equation for the effective potential in \Eq{eq:dtUk} and model parameters in the hadronic sector.} 
  \label{tab:paras}
\end{table}
%

Note that the $\sigma$-meson mass is chosen to be $m_\sigma=510$ MeV in this work; however, in the experiments the $\sigma$-meson related scalar meson $f_0(500)$ with $I^G (J^{PC})=0^{+}(0^{++})$ is located in a broad mass region of  $400-550$ MeV \cite{Patrignani:2016xqp}. Therefore, it is necessary to investigate the influence of the $\sigma$-meson mass on the QCD thermodynamics, and the relevant results are presented in \Fig{fig:mSigma1} through \Fig{fig:mSigma3} for the EoS, baryon number fluctuations, and the kurtosis of the baryon number distribution, respectively. As we discuss in \sec{sec:Num}, due to the numerical instability for the Taylor expansion of the effective potential around the physical point, we can not decrease the value of $m_\sigma$ to be below 500 MeV. With the variation of $m_\sigma$ in the interval  $510-600$ MeV, we only find a mild dependence on the $\sigma$-meson mass.

As we have mentioned above, the initial scale is chosen to be $\Lambda=1$ GeV. Making a crude estimate from the Matsubara gap as follows
\begin{align}
  \Lambda&\simeq 2\pi\,T_{\Lambda}\,,\label{eq:TLam}
\end{align}
where $T_{\Lambda}$ is the temperature, above which the UV cutoff effect becomes significant and can not be neglected. It follows from \Eq{eq:TLam} that $T_{\Lambda}\sim 160$ MeV, which is around the critical temperature of QCD phase transitions. Therefore, we have to reduce the influence of the UV cutoff on observables of interest, especially at high temperature. One way to diminish the UV cutoff effect is to introduce an appropriate modification for the effective action at the initial scale, i.e.,
\begin{align}
  \Gamma_{k=\Lambda}&\rightarrow \Gamma_{k=\Lambda}+\Delta \Gamma_{k=\Lambda}\,,\label{eq:DGam}
\end{align}
where the effects of external parameters, such as the temperature and chemical potential, can be taken into account to some degree at the second term on the r.h.s. of \Eq{eq:DGam}. For instance, one can integrate the flow of the effective action from infinity down to $\Lambda$, while with contributions of the vacuum subtracted, to wit,
\begin{align}
  \Delta \Gamma_{k=\Lambda}&=\int_{\infty}^{\Lambda}\frac{d k}{k}\left(\partial_t \Gamma_{k}\Big\vert_{T,\mu}-\partial_t \Gamma_{k}\Big\vert_{T=\mu=0}\right)\,.\label{eq:DGam2}
\end{align}
In the low energy effective model, mesons decouple and are irrelevant in the scale above the UV cutoff, thus $\Delta \Gamma_{k=\Lambda}$ in \Eq{eq:DGam2} receives contributions only from free quarks. Therefore, it follows from \Eq{eq:DGam2} that
\begin{align}
  &\Delta \Gamma_{k=\Lambda}=-\int_{\Lambda}^{\infty}d k\frac{k^3}{3\pi^2}N_c\Bigg[\frac{1}{\sqrt{1+\bar m_{l,k}^{2}}}\nonumber\\[2ex]
  &\times\Big(n_{F}(\bar m_{l,k}^{2};T,\mu_u,L,\bar L)+n_{F}(\bar m_{l,k}^{2};T,-\mu_u,\bar L,L)\nonumber\\[2ex]
  &+n_{F}(\bar m_{l,k}^{2};T,\mu_d,\bar L,L)+n_{F}(\bar m_{l,k}^{2};T,-\mu_d,\bar L,L)\Big)\nonumber\\[2ex]
  &+\frac{1}{\sqrt{1+\bar m_{s,k}^{2}}}\Big(n_{F}(\bar m_{s,k}^{2};T,\mu_s,L,\bar L)\nonumber\\[2ex]
  &+n_{F}(\bar m_{s,k}^{2};T,-\mu_s,\bar L,L)\Big)\Bigg]\,,\label{}
\end{align}
with the Polyakov-loop modified fermionic distribution function $n_{F}$ given in \Eq{eq:nFL}, and $\bar m_{l,k}^{2}=m_{l,k}^{2}/k^2$ and $\bar m_{s,k}^{2}=m_{s,k}^{2}/k^2$. When the RG scale $k$ is above $\Lambda$, the quark masses are approximated as the values of those at $k=\Lambda$ in this work, viz.
\begin{align}
  m_{l,k>\Lambda}&=m_{l,k=\Lambda}\quad\text{and}\quad m_{s,k>\Lambda}=m_{s,k=\Lambda}\,,\label{}
\end{align}
for values of $m_{l,k=\Lambda}$ and $m_{s,k=\Lambda}$, see \Fig{fig:mass}. Recently, Braun {\it et al} has proposed the concept of RG consistency, which is employed to analyze the cutoff effects, and is aimed to enhance the predictive power of low energy effective theories \cite{Braun:2018svj}.


\bibliography{ref-lib.bib}

\end{document}